\documentclass{article}
\usepackage{geometry, booktabs, subcaption}
\usepackage{titlesec}
\usepackage[T1]{fontenc}
\usepackage[utf8]{inputenc}
\usepackage{authblk}
\usepackage{amssymb,amsmath}
\usepackage{graphics}
\providecommand{\keywords}[1]{\textbf{Keywords: } #1}
\titleformat{\section}{\centering\large\scshape}{\thesection}{1em}{}
\titleformat{\subsection}{\centering\normalsize\scshape}{\thesubsection}{1em}{}

\usepackage{hyperref}
\hypersetup{
	colorlinks=true,
	linkcolor=blue,
	filecolor=blue,      
	urlcolor=blue,
	citecolor=blue
}

\usepackage{natbib}
\usepackage{graphicx} 
\graphicspath{{IMG/}}

\usepackage{setspace}
\usepackage{xr}
\usepackage{color}
\usepackage{amsmath}
\usepackage{amsfonts,dsfont,mathrsfs}
\usepackage{booktabs,multirow,array}
\usepackage{bm}
\usepackage[linesnumbered,ruled,vlined]{algorithm2e}

\makeatletter
\renewcommand{\algocf@captiontext}[2]{#1\algocf@typo. \AlCapFnt{}#2} 
\def\@algocf@capt@plain{top}
\renewcommand{\algocf@makecaption}[2]{%
	\addtolength{\hsize}{\algomargin}%
	\sbox\@tempboxa{\algocf@captiontext{#1}{#2}}%
	\ifdim\wd\@tempboxa >\hsize
	\hskip .5\algomargin%
	\parbox[t]{\hsize}{\algocf@captiontext{#1}{#2}}
	\else%
	\global\@minipagefalse%
	\hbox to\hsize{\box\@tempboxa}
	\fi%
	\addtolength{\hsize}{-\algomargin}%
}
\makeatother

\def\e{\mathrm{e}}

\def\P{\mbox{P}}

\def\d{\mathrm{d}}

\def\br{{\bf r}}

\newcommand{\bdelta}{\boldsymbol{\delta}}
\newcommand{\btheta}{\boldsymbol{\theta}}
\newcommand{\brho}{\boldsymbol{\rho}}

\newcommand{\kernel}{\mathcal K}
\newcommand{\Eval}{\mathbb{E}}
\newcommand{\Wei}{\mathcal W}
\newcommand{\partsp}{\mathcal P}
\newcommand{\partal}{\mathscr P}

\newcommand{\perm}{\lambda}
\newcommand{\uno}{\mathds{1}}

\newcommand{\iid}{\stackrel{\mbox{\scriptsize iid}}{\sim}}

\newtheorem{lemma}{Lemma}
\newtheorem{trm}{Theorem}
\newtheorem{proposition}{Proposition}

\newtheorem{defin}{\bf Definition}
\newtheorem{remark}{Remark}
\newtheorem{example}{Example}

\renewcommand{\mid}{\,|\,}
\newcommand{\eppf}[2]{\mathrm{p}_{\mathrm{#1}}^{\mathrm{(#2)}}}

\allowdisplaybreaks[3]

\begin{document}

\title{\scshape\LARGE{Bayesian nonparametric model based clustering with intractable distributions: an ABC approach}}

\author[1,2]{Beraha M.\thanks{mario.beraha@polimi.it}}
\author[3]{Corradin R.\thanks{riccardo.corradin@unimib.it}}
\affil[1]{\normalsize{Department of Mathematics, Politecnico di Milano}} 
\affil[2]{\normalsize{Department of Computer Science, Universit\`{a} degli Studi di Bologna}}
\affil[3]{\normalsize{Department of Economics, Management and Statistics,University of Milano-Bicocca}}
\date{\today}

\maketitle 

\begin{abstract}
	Bayesian nonparametric mixture models offer a rich framework for model based clustering.
	We consider the situation where the kernel of the mixture is available only up to an intractable normalizing constant. In this case, most of the commonly used Markov chain Monte Carlo (MCMC) methods are not suitable. We propose an approximate Bayesian computational (ABC) strategy, whereby we approximate the posterior to avoid the intractability of the kernel. We derive an ABC-MCMC algorithm which combines (i) the use of the predictive distribution induced by the nonparametric prior as proposal and (ii) the use of the Wasserstein distance and its connection to optimal matching problems. To overcome the sensibility with respect to the parameters of our algorithm, we further propose an adaptive strategy. We illustrate the use of the proposed algorithm with several simulation studies and an application on real data, where we cluster a population of networks, comparing its performance with standard MCMC algorithms and validating the adaptive strategy.
	
	\vspace{12pt}
	\noindent\keywords{Approximate bayesian computation; Markov chain Monte Carlo; Adaptive sampling scheme; Bayesian nonparametric; Wasserstein distance; Mixture models. }
\end{abstract}

\section{Introduction}

Approximate Bayesian computation (ABC) is a recent growing area of research, dealing with statistical problems involving intractable distributions, i.e. distributions known up to a normalizing constant or for which evaluating the probability density function is computationally prohibitive. We refer to the pioneering studies of these methodologies by mentioning the works of \citet{Rub84, Tav97, Pri99, Bea02}, among others. The principle motivation leading the introduction of these methodologies can be found in the idea of resorting to an approximate solution to the original problem. In the wide classes of possible intractable problems, we distinguish mainly between two fundamental groups. A first group can be identified by problems that are analytically intractable, in the sense that the model is not fully specified or it is not known. Within this scenario, the inferential procedures are infeasible from a calculus perspective. A second group is composed of computationally intractable problems, a common issue whenever we deal with complex models. Despite an eventual proper distribution to be sampled, providing an estimation of these models is unattainable in a feasible time.

The application of ABC methods spreads over many fields. Remarkable examples are recent usages in astronomy and cosmology \citep[e.g.][]{Cam12,Wey13}, genetics \citep[e.g.][]{Bea04,Tec15} and finance \citep[e.g.][]{Pic14,Cal14}, but not all. Many ABC methods and extensions were proposed in literature in the last decades, mainly by considering different strategies to approximate the posterior distribution, such as rejection sampler \citep[e.g.][]{Pri99,Bea02} and kernel methods \citep[e.g.][]{Bea02,Wil13} among others. These strategies can be further combined with various standard computational methods, obtaining for example ABC rejection sampler \citep[e.g.][]{Tav97, Pri99}, ABC importance sampler and sequential Monte Carlo \citep[e.g.][]{Fea12,Sis07,Sis09,Bea09}, ABC Markov chain Monte Carlo \citep[e.g.][]{Mar03,Bor07}, and ABC Variational Inference \citep[e.g.][]{Bar14,Min17}. We further refer to \citet{Kar18} for a recent and extensive review on ABC methods.

Our main objects of study along the article are latent random partitions, arising from a mixture model structure. Common techniques to deal with the estimation of these model mainly resort to Markov chain Monte Carlo (MCMC) algorithms, but they might be slow due to the computational intensity of the problem or to the model specification \citep[see, e.g.,][]{Can19}. We propose a strategy which, although approximate, is simply to implement and easily adaptable to a many prior specifications. 

We mainly focus on the class of ABC-MCMC algorithms, in the spirit of the early studies by \citet{Mar03}. 
The basic idea in ABC-MCMC is to replace the evaluation of the likelihood with the evaluation of a distance $d$ between the observed data and a synthetic dataset, generated from a surrogate model. If the true and synthetic data are close, that is if their distance is smaller than a threshold $\varepsilon$, then a Metropolis-Hastings step is performed. The key quantities to define such approximate strategy are then the choice of a metric $d$ and a threshold $\varepsilon$.

Our choice of distance $d$ is mainly motivated by the geometry of the underlying problem, that is the estimation of random partitions, and its connection with optimal transport and, therefore, the Wasserstein distance. See, e.g., \citet{Vil08} for an overview on foundations and theoretical results and \citet{peyre2019computational} for the computational aspects.
Recent attention was given in literature to combine Wasserstein distance with ABC procedures, see for example \citet{Ber19a, Ber19b}, and the related studies on coarsened posterior distributions by \citet{Mil19}. 
In this work, we propose to incorporate the usage of the Wasserstein metric as distance between true and synthetic data in a mixture model setup. 
The main advantage of using the Wasserstein metric 
is that, as a byproduct, we obtain an optimal transport map that allows us to make inference on the partition of the observed data starting from the partition of synthetic data. We further investigate the inclusion of  an adaptive strategy for $\varepsilon$, which improves the performance of the sampler while is simplifying its specification.

The paper is structured as follows: Section \ref{sec:ex_part} is a review of mixture modelling, latent random partition, intractable kernel distributions, and some results fundamental to the following sections. Section \ref{sec:ABCMCMC} introduces the ABC-MCMC sampling strategy for latent random partitions in mixture models in a general setting and discusses the use of an adaptive strategy for the rejection threshold. In Section \ref{sec:simu} we present numerical illustrations where we compare our ABC-MCMC algorithm with standard MCMC samplers based on Gibbs sampling, demonstrate the usefulness of the adaptive threshold selection strategy, and apply our algorithm to the problem of clustering a population of networks. All the routines and the algorithm used in the examples are available at \url{https://github.com/mberaha/abc_partition}.
We conclude the paper with some final comments and remarks. 

\section{Bayesian mixture models}\label{sec:ex_part}

Consider observations $\bm y_{1:n} = (y_1, \ldots, y_n)$ such that each $y_i$ belongs to a Polish space $(\mathbb Y, \mathcal{Y})$. We will always assume the Borel $\sigma$ field and skip measure-theoretic details in the following.
A possible way of specifying a mixture model is by means of a mixing distribution $\tilde p$, assuming that observations are conditionally i.i.d as follows:
\begin{equation}\label{eq:mix_lik}
	y_1, \ldots, y_n \mid \tilde p \iid \tilde f(\cdot) = \int_\Theta \kernel(\cdot \ ; \theta) \tilde p(d\theta),
\end{equation}
where $\kernel(\cdot, \cdot)$ is measurable in its two arguments, $\kernel(\cdot, \theta)$ is a probability density function for each value of $\theta \in \Theta$, and $\tilde p$ is an almost surely discrete random probability measure, i.e., $\tilde p = \sum_h w_h \delta_{\theta^*_h}$ almost surely with both the weights $w_h$'s and the atoms $\theta_h^*$'s random quantities. Note that the number of atoms in $\tilde p$ can be either finite or infinite. We further assume the distribution of the weights $w_h$'s independent of the distribution of the locations $\theta_h$'s, where the latter is usually assumed diffuse on $\Omega$.

Although our methodology is valid regardless of the specific choice of $\kernel$, we believe that it is particularly suited in cases when $\kernel$ is known up to a normalizing constant, that is:
\[
\kernel(y_i; \theta) = \frac{1}{Z_\theta} g(y_i; \theta)
\]
where $Z_\theta < +\infty$ is an intractable normalizing constant, which depends on the value of the parameters $\theta$. For further details, see Section~\ref{sec:kern}.

\begin{example}\label{ex:py1} (Pitman-Yor process mixture model) Among the possible choices of $\tilde p$, we consider $\tilde p$ distributed as a Pitman-Yor process \citep{Pit97, Ish01}. We write $\tilde p \sim PY(\vartheta, \sigma, G_0)$, where $\sigma \in [0, 1)$, $\vartheta > - \sigma$ and $G_0$ is a diffuse probability measure on $\mathbb{Y}$.
	Then, $\tilde p = \sum_{h=1}^\infty w_h \delta_{\theta^*_h}$ with $\theta^*_1, \theta^*_2, \ldots \iid G_0$ and $\{w_h\}_h$ is a sequence of weights distributed according to a two-parameter Griffiths--Engen--McCloskey distribution, i.e., $w_1 = \nu_1$, $w_h = \nu_h \prod_{j \geq h} (1 - \nu_j)$ for $h > 1$, with $\nu_h \iid Beta(1 - \sigma, \vartheta + h\sigma)$.
\end{example}

\subsection{Exchangeable random partitions and mixture models}\label{sec:exch_part}

For our purposes, it easier to think of a mixture model in terms of a latent partition and a set of cluster centers. Specifically, let $[n] = \{1, \ldots, n\}$, denote with $\bm \rho_n = A_1, \ldots, A_k$ a partition of $[n]$ (i.e. $\bigcup_j A_j = [n]$ and $A_i \cap A_j = \emptyset$ if $i \neq j$) and let $\bm \theta^* = (\theta^*_1, \ldots, \theta^*_k)$.
Writing $p(\cdot)$ for a generic density and $p(\cdot \mid \cdot)$ for a conditional density, we assume that
\begin{equation}\label{eq:lik}
	p(y_1, \ldots, y_n \mid \bm \theta^*, \bm \rho_n) = \prod_{j=1}^k \prod_{i \in A_j} \kernel(y_i; \theta^*_j),
\end{equation}
The Bayesian approach requires specifying a prior distribution for $(\bm \rho_n, \bm \theta^*)$. We assume that $\bm \rho_n$ is independent of $\bm \theta^*$ and that conditionally to the number of elements of the partition $k$ (henceforth denoted as clusters), $\theta^*_1, \ldots, \theta^*_k$ are independent and identically distributed random variables from a distribution that does not depend on $(\bm \rho_n, k)$.
For the class of distributions considered here (see below), the representations in  \eqref{eq:mix_lik} and \eqref{eq:lik}, together with prior assumptions, are indeed equivalent and, in particular, \eqref{eq:lik} can be derived from \eqref{eq:mix_lik} by marginalizing out $\tilde p$.
See, for instance, \cite{Jam09} and \cite{Pit95} and the references therein.

Let us now give further details on the class of prior distributions for $\bm \rho_n$ that we consider. Let $\partsp_{1:n}$ be the space of all possible partitions of $[n]$, denote with $\partal_{1:n}$ its discrete $\sigma$-field, then $\bm \rho_n$ is a random variable with values in $\partsp_{1:n}$, i.e. a measurable function from a generic probability space to $(\partsp_{1:n}, \partal_{1:n})$.
Given $\br_n = \{A_i\}_{i=1}^k$ a partition of $[n]$, let $\bm \rho_{n+1}$ denote the random partition of $[n+1]$. The only necessary requirement for our algorithm (cf.\ Section~\ref{sec:algo}) is that the law of the random partition allows for explicit formulas for $\P(\bm \rho_{n+1} \mid \br_n)$.
However, for simplicity and specificity, we focus here on \emph{exchangeable} random partitions.
A generic random partition $\brho_n$ of $[n]$ is said to be exchangeable if its distribution is invariant with respect to any permutation $\perm:[n] \to [n]$. 
See e.g. \citet{Kin78, Ald85} for further details. 


As shown in \citet{Pit95}, a random partition $\brho_n$ is exchangeable if and only if $\P(\brho_n = \br_n)$ depends on $\br_n = \{A_i\}_{i=1}^k$ only through the cardinalities $n_j = |A_j|$ of each set $A_j$, and there exists a symmetric function $\eppf{k}{n} (n_1, \ldots, n_k)$ named exchangeable partition probability function (EPPF) such that
\begin{equation*}
	\P(\brho_n = \br_n) = \eppf{k}{n}(\lvert A_1 \rvert, \ldots, \lvert A_k \rvert) = \eppf{k}{n}(n_1, \ldots, n_k).
\end{equation*}

This characterization helps to define of opportune distribution law for random partition objects. Clearly different choices for the EPPF function lead to different partition structures. The study of the EPPFs and their application, from a statistical perspective, is strictly related to the study of sampling scheme and predictive distribution for random partitions. In fact one can study the predictive distribution of the sampling at $n+1$, given a partition $\br_n$, by the probability of falling into a new cluster
\begin{equation}\label{eq:new_prob}
	\P (\brho_{n+1}= \{A_1, \ldots, A_k, \{n+1\}\}\mid \br_n) = \frac{\eppf{k+1}{n+1}(n_1, \ldots, n_k, 1)}{\eppf{k}{n}(n_1, \ldots, n_k)},
\end{equation}
and the probability of falling in the $j$-th already observed one as
\begin{equation}\label{eq:old_prob}
	\P (\brho_{n+1}= \{A_1, \ldots, A_j \cup \{n+1\}, \ldots, A_k\}\mid \br_n) = \frac{\eppf{k}{n}(n_1, \ldots, n_j + 1, \ldots, n_k)}{\eppf{k}{n}(n_1, \ldots, n_k)}.
\end{equation}

Assuming that $\{\theta_h\}_{h\geq 1}$ is an exchangeable sequence from an a.s. discrete random probability measure $\tilde p$ is tantamount to say that there exists a latent random partition in the sequence describing possible ties in the sequence, and further we have that for any $n\in \mathbb N$ the latent partition in $\theta_1, \dots, \theta_n$ can be derived as 
\[
\eppf{k}{n}(n_1, \ldots, n_k)= \int_{\mathbb Y^k} \Eval\left[\prod_{j=1}^k \tilde p^{n_j}(\mathrm{d}\theta_j^*)\right].
\]

\begin{example}\label{ex:py2} (Pitman-Yor process mixture model (continued)) the EPPF of a PY process can be explicitly characterized as
	\[
	\eppf{k}{n}(n_1, \ldots, n_k) = \frac{\prod_{j=1}^{k-1} (\vartheta + j\sigma)}{(\vartheta + 1)_{n-1}} \prod_{j=1}^k (1 - \sigma)_{n_j -1},
	\]
	where $(x)_n = x (x +1) \cdots (x + n -1)$ denotes the Pochhammer symbol. Moreover it is straightforward to derive simpler expressions for the probabilities in \eqref{eq:new_prob}-\eqref{eq:old_prob}
	\[
	\P (\brho_{n+1}= \{B_1, \ldots, B_k, \{n+1\}\}\mid \br_n) = \frac{\vartheta + k \sigma}{\vartheta + n},
	\] 
	and
	\[
	\P (\brho_{n+1}= \{B_1, \ldots, B_j \cup \{n+1\}, \ldots, B_k\}\mid \br_n) = \frac{n_j - \sigma}{\vartheta + n}.
	\]
\end{example}

\subsection{Dealing with intractable kernel density functions}\label{sec:kern}

When the kernel $\kernel$ is known up to a normalizing constant, traditionally employed MCMC algorithms become impractical. Considering for instance Neal's Algorithm 2 \citep{Nea00} and its generalizations, which represents one of the cornerstone strategies to deal with mixture models with almost surely discrete random probability measures as mixing distributions, the following updates are required:
\begin{itemize}
	\item[(A)]  for the cluster-specific parameters $\theta_h^*$, sample each $\theta_h^*$ from the full conditional
	\[
	p(\theta_h^* \mid \cdots) \propto \prod_{i \in A_h} \kernel(y_i; \theta_h^*) G_0(\theta_h^*) = \left(\frac{1}{Z_{\theta_h}}\right)^{|A_h|} \prod_{i \in B_h} g(y_i; \theta_h^*) G_0(\theta_h^*)
	\]
	where we write $p(\theta_h^* \mid \cdots)$ for the full conditional distribution of $\theta_h^*$ conditioned on all the other parameters and the data, and $G_0$ is distribution possibly diffuse over $\Theta$.
	\item[(B)] For the latent partition $\bm \rho_n$, update the cluster allocation of each observation sampling from
	\[
	\P(i \in A_h \mid \cdots) \propto \begin{cases}
	\eppf{k}{n}(n^{-i}_1, \ldots, n^{-i}_h + 1, \ldots n^{-i}_k) g(y_i; \theta_h^*) /Z_{\theta_h} &:h = {1,\dots, k}  \\
	\eppf{k}{n}(n^{-i}_1, \ldots n^{-i}_k, 1) \int_{\Theta } g(y_i; \theta) / Z_{\theta} p(d\theta)&: h = k + 1
	\end{cases}
	\]
	where the superscript $-i$ means that the $i$-th observation has been removed from the calculations.
\end{itemize}

Both updates present non-trivial challenges. Step (A) above requires sampling from a so-called doubly intractable distribution. Assuming that a perfect simulation algorithm from $\kernel$ is available, sampling from $p(\theta_h \mid \bm \cdots)$ can be performed through an exchange algorithm as the one in \cite{moller2006efficient}. However, as pointed out in \cite{murray2006mcmc} the exchange algorithm can lead to low acceptance rates and a better solution would be to employ a sequence of tempered transitions, which still requires nontrivial implementations and fine-tuning.
Step (B) involves a distribution over the integers $\{1, \ldots, k+1\}$.
The probability associated to $k+1$ involves an integral, but this can be overcome by using for instance Neal's Algorithm 8. Hence, for the sake of the argument, let us ignore the last term.
Usually, one simply computes the unnormalized probabilities, normalizes them and samples from the resulting discrete probability distribution.
However in this case, each term also contains $Z_{\theta_h}$ which is unknown, so that this simple strategy is not possible.
One could instead employ a Metropolis-Hastings step with a proposal over $\{1, \ldots, k+1\}$, which would require again the use of some form of the exchange algorithm to get rid of the ratios of normalizing constants. In summary, the presence of an intractable normalizing constant in $\kernel$ severely impacts the feasibility of commonly used MCMC algorithms for mixture models and presents a major bottleneck for efficiency.

\section{ABC-MCMC for random partitions}\label{sec:ABCMCMC}

By applying Bayes' theorem, the posterior of the partition $\bm \rho_n$ given data $\bm y$ can be written as
\begin{equation}\label{eq:posterior}
	\pi(\brho_n \mid \bm y_{1:n}) = \frac{p(\brho_n)p(\bm y_{1:n} \mid \brho_n)}{ \sum_{\brho_n \in \partsp_{1:n}}p(\d \brho_n)p(\bm y_{1:n}\mid \brho_n)},
\end{equation}
where 
\[
p(\bm y_{1:n} \mid \rho_n) = \prod_{j=1}^k \int \prod_{i \in A_j} \kernel(y_i; \theta) G_0(\d \theta) = \prod_{j=1}^k \int \prod_{i \in A_j} \frac{g(y_i; \theta)}{Z_\theta} G_0(\d \theta).
\]

As standard in the ABC framework, we consider an $\varepsilon$-approximation $\pi_\varepsilon$ of \eqref{eq:posterior}, by introducing  a set of synthetic data $\bm s_{1:n}\in \mathbb{Y}^n$. Let  $d: \mathbb{Y}^n \times \mathbb{Y}^n \rightarrow [0, +\infty)$ be a metric (we will discuss specific choices later), then
\begin{equation}\label{eq:appr_post}
	\pi_\varepsilon(\brho_n \mid \bm y_{1:n}) = \frac{ p(\brho_n) \int_{\mathbb{Y}^n}\uno_{[d(\bm y_{1:n}, \bm s_{1:n}) < \epsilon]} p(d \bm s_{1:n}\mid \brho_n)}{ \sum_{\brho_n \in \partsp_{1:n}}p(\d \brho_n) \int_{\mathbb{Y}^n}\uno_{[d(\bm y_{1:n}, \bm s_{1:n}) < \epsilon]} p(d \bm s_{1:n}\mid \brho_n)}.
\end{equation} 
Starting from \eqref{eq:appr_post}, a basic acceptance-rejection ABC algorithm can be straightforwardly derived considering the following steps:
\begin{enumerate}
	\item Sample a partition $\widetilde{\bm \rho}_n = \{A_j\}_{j=1}^k$ from the prior.
	\item Conditionally on $\widetilde{\bm \rho}_n$, sample $\theta^*_j \iid G_0$ $j=1, \ldots, k$ and $\{s_i\}_{i \in A_j}\mid \theta^*_j \iid \kernel(\cdot; \theta^*_j)$.
	\item Accept $\widetilde{\bm \rho}_n$ if $d(\bm s_{1:n}, \bm y_{1:n}) < \varepsilon$.
\end{enumerate}
Although it is not the one we will employ (see Section~\ref{sec:ABCMCMC}), it is instructive to report it here for the discussion below.
First, we note that the distance $d(\cdot, \cdot)$ has not been specificed yet. Traditionally, ABC algorithms employed statistics $\nu: \mathbb Y^n \rightarrow \mathbb{R}^d$ and considered $d(\bm y_{1:n}, \bm s_{1:n}) = \| \nu(\bm y_{1:n}) - \nu(\bm s_{1:n}) \|$. For instance $\nu$ could compute the mean and variance of $\bm y_{1:n}$. The use of summary statistics simplifies the computations as it allows for great dimensionality reduction but it also causes a loss of information. Further, the choice of which summary statistics to use is not obvious \citep{Fea12}.
More recently, the use of statistical distances to compare the empirical distributions of $\bm y_{1:n}$ and $\bm s_{1:n}$ has been proposed, to overcome the issues related to summarization. See, for instance, \cite{Dro21} and the references therein.
Moreover, another issue is evident when inspecting the output of the acceptance-rejection algorithm. In fact, the partition $\widetilde{\bm \rho}_n$ accepted at step 3. above, is referred to $\bm s_{1:n}$ and provides little information about the clustering of the observations $\bm y_{1:n}$.
In the following, we show how to overcome both issues by a suitable choice of the distance, namely, the Wasserstein distance.

Given two measures $\mu_1, \mu_2$ over $\mathbb{Y}$ with finite $q$-th moment and a cost function $c: \mathbb{Y} \times \mathbb{Y} \rightarrow [0, +\infty)$, assumed convex in the following, the Wasserstein distance of order $q$ is defined as
\begin{equation}\label{eq:WEI}
	\Wei_q(\mu_1, \mu_2):=\left\{ \inf_{\gamma \in \Gamma(\mu_1, \mu_2) }  \int_{\mathbb{Y} \times \mathbb{Y}} c(x_1, x_2) \mathrm{d}\gamma(x_1, x_2) \right\}^{\frac{1}{q}}
\end{equation}
where $\Gamma(\mu, \nu)$ denotes the Radon space of all measures defined on $\mathbb{Y} \times \mathbb{Y}$ with marginals $\mu_1$ and $\mu_2$.
Letting $\mu_1 = n^{-1} \sum \delta_{y_i}$ and $\mu_2=n^{-1} \sum \delta_{s_i}$, we can use $\Wei_q$, with a suitable choice of cost function, to compare $\bm y_{1:n}$ and $\bm s_{1:n}$. We will write $\Wei_q(\bm y_{1:n}, \bm s_{1:n})$ to make this explicit.
This is the case of the Wasserstein-ABC algorithm in \cite{Ber19b}, where the authors propose to use the Wasserstein distance principally
to avoid the choice of statistics for the ABC-SMC scheme. See also e.g. \cite{bassetti2006minimum} and \cite{Ber19a} for further uses of the Wasserstein distance in the statistical framework.

In this work, the Wasserstein distance is not only useful to avoid summarization, but is also the key ingredient that allows to make inference on the partition of $\bm y_{1:n}$ starting from the partition of $\bm s_{1:n}$, $\widetilde{\bm \rho}_n$. First, we note that since $\mu_1$ and $\mu_2$ are always discrete measures, \eqref{eq:WEI} reduces to
\begin{equation}\label{eq:WEI2}
	\Wei_q(\bm y_{1:n}, \bm s_{1:n}):=\left\{ \min_{P \in M_{n\times n} } \sum_{i=1}^n \sum_{j=1}^n c(y_i,  s_j)^q P_{i,j} \right\}^{\frac{1}{q}} = \left\{ \min_{P \in M_{n\times n} } <C^{(q)}, P> \right\}^{\frac{1}{q}}
\end{equation}
where, referring to an optimal transport notation, $C^{(q)}$ denotes the cost matrix of order $q$, with $i,j$-th element $C^{(q)}_{i,j} = c(y_i,  s_j)^q$ and $P$ denotes the transport matrix. Observe also how infimum in \eqref{eq:WEI} has been replaced with a minimumn in \eqref{eq:WEI2}.
The following proposition clarifies the connection between the Wasserstein distance and permutation matrices
\begin{proposition}\label{prop:opt}
	Let $\mu = \sum_{i=1}^n a_i \delta_{x_i}$ and $\nu = \sum_{i=1}^m b_i \delta_{y_i}$.
	If $m=2$ and $\bm a = \bm b$, $a_i = b_i = 1 / n$ for all $i\in \{1,\dots, n\}$, then there exist an optimal solution to problem  \eqref{eq:WEI2} $P^* = P_{\lambda^*}$, which is a permutation matrix associated to an optimal permutation $\lambda^*$ in the class of permutations of $n$ elements.
\end{proposition}

We refer to Proposition 2.1 of \citet{peyre2019computational} for a detailed proof of Proposition \ref{prop:opt}. Hence, by computing the Wasserstein distance between $\bm y_{1:n}$ and $\bm s_{1:n}$, we are also matching the partition $\widetilde \brho_n$ of $\bm s_{1:n}$ to a corresponding partition $\brho_n$  of $\bm y_{1:n}$, by considering $\brho_n = \perm^*(\widetilde \brho_n)$ .
This result is remarkable as it allows to find in a polynomial time a solution 
to the assignment problem, while the space of all permutation of $n$ objects has size $n!$.
Hence, when the distance $ \Wei_q(\bm y_{1:n}, \bm s_{1:n})$ is less than the threshold $\varepsilon$, we accept $\perm^*(\widetilde \brho_n)$ as realization from $\pi_\varepsilon(\brho_n \mid \bm y_{1:n})$. We further remark that such rearrangement is legit in force of the exchangeability of the observe data.

\subsection{Computation of the Wasserstein distance}

When the data are univariate, computing the Wasserstein distance between $\bm y_{1:n}$ and $\bm s_{1:n}$ and the related optimal permutation can be efficiently done, as the minimization problem is available in close form, as in the following remark \citep[remark 2.30 in][]{peyre2019computational}.

\begin{remark}
	For measures $\mu, \nu$ on $\mathbb{R}$, denote with $F_\mu$ ($F_\nu$) the cumulative distribution function of $\mu$ ($\nu$) from $\mathbb{R}$ to $[0, 1]$, defined as
	\[
	F_\mu(x) = \int_{-\infty}^x d\mu \quad \text{for all } x 
	\] 
	and its pseudoinverse 
	\[
	F^{-1}_\mu(x) = \min_z \{z \in \mathbb{R} \cup \{-\infty\}: F_\mu(z) \geq x \}.
	\]
	Then for any $q \geq 1$ one has
	\[
	\Wei_q(\mu, \nu)^q = \int_0^1 \lvert F_\mu^{-1}(x) - F_\nu^{-1}(x) \rvert^q dx
	\]
\end{remark}
Letting $\mu=\hat \mu$ and $\nu=\hat \nu$, it is apparent that the optimal solution is given by sorting both the vectors $\bm y_{1:n}$ and $\bm s_{1:n}$. The computational cost of solving the problem in an optimal way is of order $n \log n$. In the multivariate setting \eqref{eq:WEI2} can be solved exactly using the Hungarian algorithm, which has a cost of order $n^3$.
This cost can become prohibitive for large sample sizes, but we can resort to an approximation of the Wasserstein distance which yields a great saving in terms of computational time.

Let $\epsilon \geq 0$ a real valued regularization term. By introducing an entropic regularization factor in \eqref{eq:WEI} we obtain the so called Sinkhorn distance
\[
W_q^{\epsilon}(\mu, \nu) = \min_{\gamma \in \Gamma(\mu, \nu)}\int \lvert \lvert  x- s \rvert \rvert_q d\gamma(x, s) + \epsilon KL(\gamma \vert \vert \mu \otimes \nu)
\]
where $KL$ is the Kullback Leibler divergence 
and $\mu \otimes \nu$ denotes the product measure.

In the case of discrete measures, \citet{Cut13} showed that solution to the Sinkhorn distance can be computed by an iterative algorithm, which requires a cost of $n^2$ per iteration and can be shown to converge in $O(\epsilon^{-2})$ iterations, up to a logarithmic factor. Moreover, one has that the Sinkhorn distance converges to the regular Wasserstein distance as $\epsilon \rightarrow 0$. More recently, \citet{altschuler2017near} proposed a greedy variant of the original Sinkhorn algorithm, which runs in a nearly linear time.
Nonetheless, both these algorithm require the computation of the full pairwise cost matrix $C$, which is still $O(n^2)$.

\subsection{ABC-MCMC approach for random partitions}\label{sec:algo}


The main problem we face when sampling from $\pi_\varepsilon$ is that the size of the partitions' space $\mathcal P_{1:n}$ escalates quickly as $n$ increases (it is equal to the $n$--th Bell number), which makes the acceptance-rejection ABC algorithm useless in practical applications.
To overcome this issue, below we discuss an ABC-MCMC sampling scheme based on the predictive distribution of the prior for the random partition.


We remark that, the predictive distribution for the $n+1$ latent parameter $\theta_{n+1}$ given $\theta_1, \dots, \theta_n$, and assuming the $\theta_i$'s an exchangeable sample from $\tilde p$, is given by 
\begin{equation}\label{eq:pred_theta}
	\P(\theta_{n+1}\in \d t \mid \theta_1, \dots, \theta_n) = \frac{\eppf{n+1}{k+1}(n_1, \dots, n_k, 1)}{\eppf{n}{k}(n_1, \dots, n_k)}G_0(\d t) + \sum_{j=1}^k \frac{\eppf{n+1}{k}(n_1, \dots,n_j + 1, \dots n_k)}{\eppf{n}{k}(n_1, \dots, n_k)}\delta_{\theta_j^*}(\d t)
\end{equation}
where $\theta_1^*, \dots, \theta_k^*$ denote the unique values in $\theta_1, \dots, \theta_n$ and  $n_1, \dots, n_k$ their frequencies 
Observe that the predictive distribution is a convex combination of the prior guess, expressed in terms of $G_0$, and the empirical information of the previous values of the latent parameters, driven by the EPPFs' ratios.

We can exploit the chain rule to produce a sample $n$ step further from the current state, obtaining 
\begin{equation}\label{eq:chain_rule}
	\P(\btheta_{n+1,2n}\mid\btheta_{1:n}) = \P(\theta_{n+1}\mid\btheta_{1:n}) \P(\theta_{n+2}\mid\btheta_{1:n}, \theta_{n+1})\dots\P(\theta_{2n}\mid\btheta_{1:n},\theta_{n+1},\dots, \theta_{2n-1}).
\end{equation}
Since at each step of the chain rule we are using the predictive distribution in \eqref{eq:pred_theta}, the resulting $\btheta^\prime_{1:n} = \btheta_{n+1,2n}$, is a combination of the prior guess and the empirical information of $\btheta_{1:n}$. 
Thanks to the fact that $\theta_i \mid \tilde p \sim \tilde p$, we can think on $\btheta^\prime_{1:n}$ as effective sample from $\tilde p$, with latent partition $\brho^\prime_n$ here termed raw candidate. We can then sample a set of synthetic data $\bm s_{1:n}$ conditionally on $\btheta^\prime_{1:n}$, with the generic $S_i \sim \kernel(s_i; \theta_i^\prime)$ for all $i\in\{1,\dots,n\}$.

Once we have produced a set of synthetic data, we evaluate its distance from the observed data $\bm y_{1:n}$ via Wasserstein metric, by solving $\Wei_q(\bm y_{1:n}, \bm s_{1:n})$. The evaluation of such distance is producing also an optimal permutation of the synthetic data, according to \eqref{eq:WEI2}, which indeed is permuting also the raw candidate of the latent partition $\brho^\prime_n$, producing an optimal candidate denoted by $\brho_n^{\prime\prime}$. Whenever $\Wei_q(\bm y_{1:n}, \bm s_{1:n})$ is smaller than a threshold $\varepsilon$, we can perform a Metropolis-Hastings step to update the state of the latent partition, or stay on the current value.

We notice that by proposing a partition according to \eqref{eq:pred_theta} and \eqref{eq:chain_rule}, we are always accepting $\brho_n^{\prime\prime}$ as far as the Wasserstein distance is smaller than $\varepsilon$, indeed denoting by $q(\btheta_{1:n} \to \btheta^{\prime\prime}_{1:n}) = \P(\btheta^{\prime\prime}_{1:n} \mid \btheta_{1:n})$ the proposal distribution, with $\btheta_{1:n}^{\prime\prime}$ the optimal permuted version of $\btheta_{1:n}^\prime$, the acceptance rate of the Metropolis-Hasting step is equal to
\begin{equation*}
	\alpha(\brho_n^{\prime\prime}, \brho_n) = 1 \wedge \frac{\P (\btheta_{1:n})  q(\btheta_{1:n} \to \btheta^{\prime\prime}_{1:n} )}{\P (\btheta_{1:n}^{\prime\prime})q(\btheta^{\prime\prime}_{1:n} \to \btheta_{1:n})} =  1 \wedge  \frac{\P (\btheta_{1:n})\P (\btheta^{\prime\prime}_{1:n} ,\btheta_{1:n}) \P(\btheta^{\prime\prime}_{1:n})}{\P (\btheta^{\prime\prime}_{1:n})\P (\btheta_{1:n}, \btheta^{\prime\prime}_{1:n})\P (\btheta_{1:n})} =1,
\end{equation*}
where $\P (\btheta^{\prime\prime}_{1:n} ,\btheta_{1:n}) = \P (\btheta_{1:n} ,\btheta_{1:n}^{\prime\prime}) $ in force of the exchangeability of the latent parameters. Such behavior is caused by the usage of the predictive distribution as proposal distribution, which is reminding the studies of \cite{Cla20}. 

\SetNlSty{textbf}{[}{]}
\begin{algorithm}[h!]
	\DontPrintSemicolon
	\textbf{input}{ a set of data $\bm y_{1:n}$, a threshold $\varepsilon$,  and possibly hyperparameters for $\kernel(\cdot; \theta)$};\\
	\textbf{set} admissible initial values for $\btheta^{(0)}_{1:n}$;\\
	\For{$r = 1,\dots,R$}    
	{ 
		\SetKwProg{Repeat}{repeat}{}{}
		\Repeat{}{
			\textbf{propose} a move from $\btheta^{(r-1)}_{1:n}$ to $\btheta^\prime_{1:n}$ according to a transition kernel $q(\btheta^{(r-1)}_{1:n} \to \btheta^{\prime}_{1:n}),$ with related partition $\brho_n^\prime$;\\ 
			\textbf{sample} $\bm s_{1:n}\mid \btheta^\prime_{1:n}$ vector of synthetic data, where $S_i \sim \kernel(\cdot, \theta^\prime_i)$;
		}
		\textbf{until} $\Wei_q(\bm y_{1:n}, \bm s_{1:n}) < \varepsilon$;\\
		
		\textbf{accept} $\brho_n^{\prime\prime}$, the permuted version of $\brho_n^{\prime}$, as realization from $\pi_\varepsilon(\brho_n \mid \bm y_{1:n})$;
	}
	\textbf{end}
	\caption{\label{algo:ABCMCMCpart}ABC-MCMC for latent random partitions}
\end{algorithm}


An implementation of the previous strategy is reported in Algorithm \ref{algo:ABCMCMCpart}. We can easily prove that Algorithm~\ref{algo:ABCMCMCpart} is effectively producing $R$ realizations from a Markov chain which has invariant distribution corresponding to the $\varepsilon$-approximation of the posterior distribution. We synthesize in the following Lemma the convergence of Algorithm~\ref{algo:ABCMCMCpart}. 

\begin{lemma}\label{theo:inv}
	Assume $\{\brho_{n,1}, \brho_{n,2}, \ldots\}$ be a sample from an ABC-MCMC scheme according to algorithm \ref{algo:ABCMCMCpart}, with proposal $q(\brho_{n} \to \brho_{n}^\prime)$ described in \eqref{eq:pred_theta} and \eqref{eq:chain_rule}. Then the produced chain has invariant distribution $\pi_\varepsilon(\brho_{1:n} \mid \bm y_{1:n})$.
\end{lemma}

The proof of Lemma~\ref{theo:inv} is trivial, as consequence of \citet{Mar03}. For the sake of completeness, and to help the understanding of the proof of Theorem~\ref{th:adapt}, we report in Appendix~\ref{app:proof_inv} a proof of Lemma~\ref{theo:inv}. The presented strategy could be a first simple approach to perform approximate inference of latent random partition, nevertheless we can relax the assumption of a fixed threshold $\varepsilon$ along the chain.

\subsection{An adaptive strategy for $\varepsilon$}

The threshold $\varepsilon$ has a strong impact on the computational time and the quality of the results of Algorithm~\ref{algo:ABCMCMCpart}, 
with small $\varepsilon$ leading to a bad mixing (it is hard to accept a proposed value) and large $\epsilon$ providing a rough approximation of the true posterior. Choosing a suitable threshold seems an essential task, but as remarked by \cite{Vih20} threshold selection \citep[see e.g.][]{Bea02, Weg09} may be not suitable in a MCMC regime with weak informative prior. Instead of a fixed $\varepsilon$, a possible strategy is to consider a sequence $\{\varepsilon_l\}_{l\geq 1}$, which allows for larger thresholds in the early phase of the chain, leading to a larger acceptance rate in the early phase of the algorithm. 

\SetNlSty{textbf}{[}{]}
\begin{algorithm}[h!]
	\DontPrintSemicolon
	\textbf{input}{ a set of data $\bm y_{1:n}$, a threshold $\varepsilon$, and possibly hyperparameters for $\kernel(\cdot; \theta)$};\\
	\textbf{set} admissible initial values for $\btheta^{(0)}_{1:n}$, set $l = 1$;\\
	\For{$r=1,\dots, R$}   
	{ 
		\SetKwProg{Repeat}{repeat}{}{}
		\Repeat{}{
			\textbf{propose} a move from $\btheta^{(r-1)}_{1:n}$ to $\btheta^\prime_{1:n}$ according to a transition kernel $q(\btheta^{(r-1)}_{1:n} \to \btheta^{\prime}_{1:n}),$ with related partition $\brho_n^\prime$;\\ 
			\textbf{sample} $\bm s_{1:n}\mid \btheta^\prime_{1:n}$ vector of synthetic data, where $S_i \sim \kernel(\cdot, \theta^\prime_i)$;\\
			\textbf{update} $\varepsilon_l$ and set $l = l + 1$;\\
		}
		\textbf{until} $\Wei_p(\bm y_{1:n}, \bm s_{1:n}) \leq \varepsilon_\ell$;\\
		\textbf{accept} $\brho_n^{\prime\prime}$, the permuted version of $\brho_n^{\prime}$, as realization from $\pi_{\varepsilon_l}(\brho_n \mid \bm y_{1:n})$;
	}
	\textbf{end}
	\caption{\label{algo:adaptABC}adaptive ABC-MCMC for latent random partitions}
\end{algorithm}

Algorithm \ref{algo:adaptABC} describes an implementation of the adaptive strategy. We remark that while we are sampling $R$ values from the approximate posterior distribution, the update of the threshold can be done also when we are rejecting the proposed values. 

By assuming that the sequence $\{\varepsilon_l\}_{l\geq 1}$ is converging to a fixed threshold $\varepsilon^*$, we are able to characterize the limit behavior of the target distribution, showing that such case has invariant distribution corresponding to the $\varepsilon^*$-approximation of the posterior distribution $\pi_{\varepsilon^*}(\brho_{n} \mid \bm y_{1:n})$, as stated in the following Theorem. 
\begin{trm}\label{th:adapt}
	Let $\{\varepsilon_l\}_{l\geq 1}$ be $\mathbb R^+$-valued sequence of elements, such that $\lim_{l\to +\infty} \lvert \varepsilon_l - \varepsilon^*\rvert = 0$. Let $\left\{\brho_{n}^{(1)}, \brho_{n}^{(2)}, \ldots\right\}$ be a sample from an ABC-MCMC scheme according to Algorithm \ref{algo:adaptABC}, with proposal $q(\brho_{n} \to \brho_{n}^\prime)$ according to \eqref{eq:pred_theta} and \eqref{eq:chain_rule}. Let $p(w)$ denotes the density function of $\Wei_q(\bm y_{1:n}, \bm s_{1:n})$, where $\bm s_{1:n}$ denotes the $l$-th synthetic sample, and assume $0 < p(w) < M$ for all $l$. Then, for $l \to +\infty$, we have that $\pi_{\varepsilon^*}(\brho_{n} \mid \bm y_{1:n})$ is a.s. the invariant distribution of the chain. 
\end{trm}
To prove Theorem~\ref{th:adapt} we exploit the continuity of $\Wei_p(\bm y_{1:n}, \bm s_{1:n})$, and the convergence of $\{\varepsilon_t\}_{t\in T}$ to its limit. A detailed proof is reported in Appendix~\ref{app:ad_abc_mcmc}.


The sequence of thresholds $\{\varepsilon_l\}_{l\geq 1}$ can be specified in many ways. A possible strategy is to consider the generic $l$-th value $\epsilon_l$ as function of a large initial value $\varepsilon_0$, a small limit value $\varepsilon^*$, and a function of the observed values of the distance until the $l$-th step, denote by $\{d_1, \dots, d_{l-1}\}$. We can for example consider a convex combination of the previous quantities as 

\begin{equation}\label{eq:ada_fun}
	\varepsilon_{l} = w_{1,l} \varepsilon_0 + w_{2,l} g(d_1, \dots, d_{l-1}) + (1 - w_{1,l} - w_{2,l}) \varepsilon^*,
\end{equation}
with $0 \leq w_{1,l} + w_{2,l} \leq 1$, the weights $w_{1,l},w_{2,l}$ are driven the smoothness of passing from $\varepsilon_0$ to $\varepsilon^*$, and $g(\cdot)$ is an opportune function of the previously observed values of the Wasserstein distance. See Section \ref{sec:simu} for a specific choice of $g(\cdot)$, $w_{1,l}$ and $w_{2,l}$. The limit value $\varepsilon^*$ can be then choose via different strategies: one can resort, for example, to the Algorithm 3 of \cite{Vih20} for the definition of a pre--processing strategy with an ideal acceptance rate. In some scenarios we found that an adaptive scheme over entire chain leaded to slightly better numerical results (cf. Section \ref{sec:simu}). Within this strategy, we waive controlling the degree of approximation of the target distribution to increase the flexibility of the sampling strategy, since the threshold might increase or decrease over the entire sampled chain.

\section{Numerical illustrations}\label{sec:simu}

Here we investigate the performance of the ABC-MCMC strategy, possibly including the adaptation on the threshold, with a standard P\'olya urn based approach, resorting to a marginal sampler \citep{Esc88,Esc95}. Once we have ran the algorithms, we estimate the optimal latent partition of the data resorting to a decision theory approach based on the \textit{variation of information} loss function \citep{Wad18, Ras18}. We further compare the estimated latent partition with the true partition by measuring the normalized variation of information, as
\[
\mathrm{VI}(\bm r_1, \bm r_2) = \frac{1}{\log n} (\mathrm{H}(\bm r_1) + \mathrm{H}(\bm r_2) - 2 \mathrm{I}(\bm r_1, \bm r_2))
\]
where $\bm r_\ell = \{A_{1,\ell}, \dots, A_{k_\ell, \ell}\}, \,\ell =1,2$, $\mathrm{H}(\bm r) = \sum_{j>1}^k p_j(\bm r) \log p_j(\bm r)$ represents the entropy associated to the partition $\bm r$, while $\mathrm{I}(\bm r_1, \bm r_2) = \sum_{i=1}^{k_1}\sum_{j=1}^{k_2} p_{ij}(\bm r_1, \bm r_2)\log [p_{ij}(\bm r_1, \bm r_2) / (p_i(\bm r_1)p_j(\bm r_2))]$ denotes the mutual information of $\bm r_1$ and $\bm r_2$, with $p_j(\bm r_\ell) = \lvert A_{j,\ell}\rvert/n$, $p_{ij}(\bm r_1, \bm r_2) = \lvert A_{i,1} \cap A_{j,2}\rvert /n$, and $k$, $k_1$, $k_2$ denote the cardinality of $\bm r$, $\bm r_1$, $\bm r_2$ respectively. Lower values of the variation of information indicate that $\bm r_1$ and $\bm r_2$ are  close to each others. We measure the mixing of the chain by considering two functionals of the visited partitions $\{\bm r^{(j)}\}_{j\geq 1}$, namely the number of clusters $k(\bm r^{(r)}) := \lvert \bm r^{(r)}\rvert$ and the entropy $\mathrm H(\bm r^{(j)})$.

In the following, we assume a Pitman-Yor mixing measure, with strength parameter $\vartheta = 1$, discount parameter $\sigma = 0.2$. The base measure $G_0$ will be specified depending on the specific example. 
When considering the adaptive ABC-MCMC, the optimal acceptance rate is set equal to $\alpha^* = 0.1$, according to \citet{Vih20}. For the adaptive ABC-MCMC with the adaptation stopped after the burn-in phase (ABCad1 in the following examples), we consider a specification of~\eqref{eq:ada_fun} with $ w_{1,l} = \e^{-(l - l_{\text{burn}}) / 1000}$, $ w_{2,l} = (1 - w_{1,l})\e^{-(l - l_{\text{burn}}) / 1000}$, and the function $g(\cdot)$ equals the $0.1$ rolling quantile of the last $100$ observed distances, where $l_{\text{burn}}$ stands for the global number of trials during the burn-in phase. The MCMC chains are sampled for $20\,000$ iterations, discarding the first $10\,000$ as burn-in. 


\subsection{Mixture of univariate Gaussian distributions}

We simulate $n$ data from a unbalanced mixture of Gaussian distributions $f_0$, with 
\[
f_0(x) = 0.75 \phi(x;-3, 1) + 0.25 \phi(x;3, 1)
\]
where $\phi(\cdot; \mu, \sigma^2)$ denotes the probability density function of a Gaussian distribution with expectation $\mu$ and variance $\sigma^2$. We consider across the study different sample sizes, with $n \in \{100, 250\}$. We set as prior model a Gaussian mixture with Pitman-Yor process mixing measure. The base measure $G_0(\mu, \sigma^2)$ is the Normal-Inverse-Gamma distribution, i.e. $\sigma^2 \sim \mathrm{IG}(2, 2)$ and $\mu\mid\sigma^2 \sim \mathrm{N}(0,  2 \sigma^2)$.

We use five different sampling strategies to face posterior inference: a non-adaptive ABC-MCMC where $\varepsilon = \sqrt{n \log n}$ (ABC1), another non-adaptive ABC-MCMC with a slightly smaller threshold $\varepsilon = 0.9 \sqrt{n \log n}$ (ABC2), the adaptive ABC-MCMC with adaptation only during the burn-in phase (ABCad1), the adaptive ABC-MCMC with adaptation through all the chain (ABCad2), and the Algorithm 2 in \cite{Nea00}, which is a popular example of ``marginal'' MCMC algorithm for BNP models.

%
%
%
%

\begin{figure}[h!]
	\includegraphics[width = 0.99\textwidth]{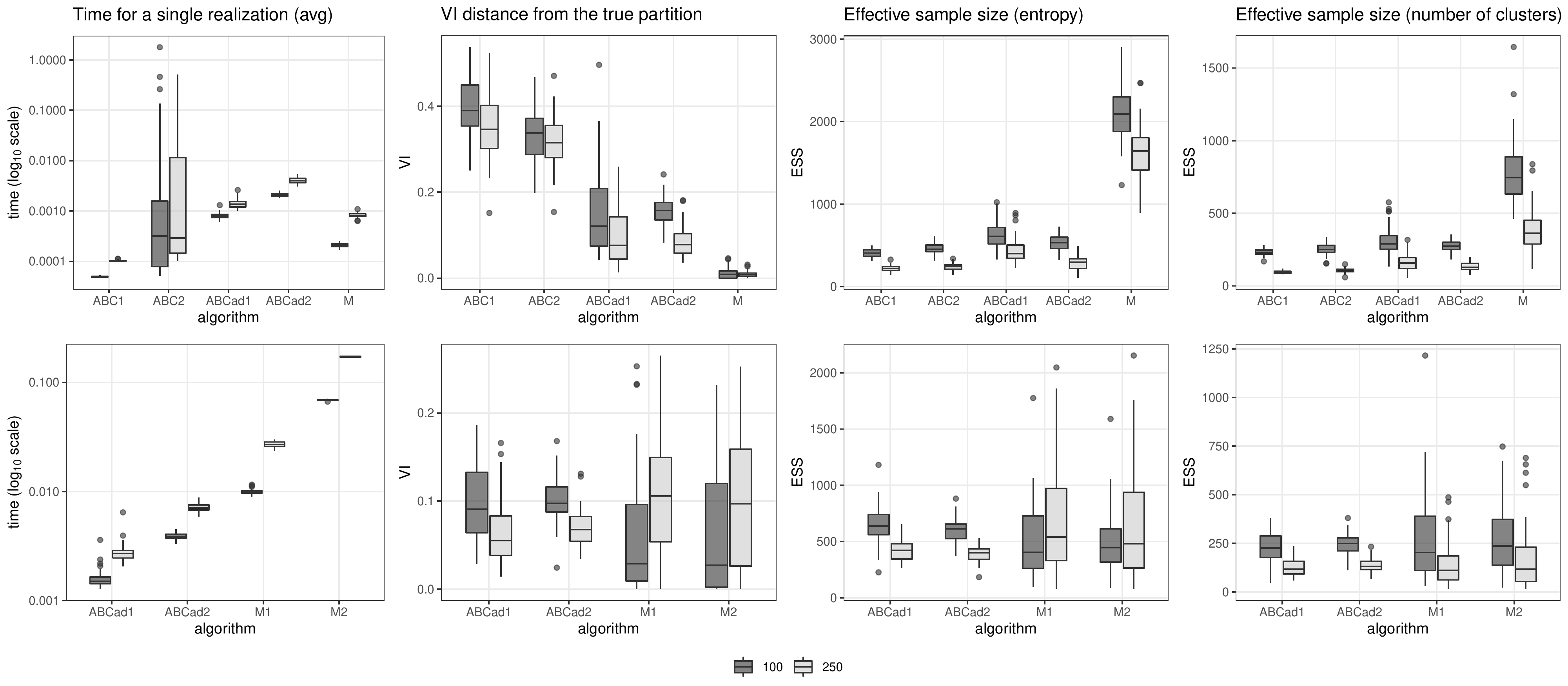}
	\caption{\label{fig:scen1} Simulation summaries. Different sample sizes $n \in \{200, 250\}$ (dark gray and light gray respectively). The results are averaged over $100$ of replications. Left to right panels: time for a single iteration, on a $\log_{10}$ scale; VI distance between the point estimate of the latent partition and the true partition; effective sample size of the entropy; effective sample size of the number of clusters. Top to bottom: Gaussian data and g-and-k data. Different sampling strategies: ABC-MCMC algorithm with large threshold (ABC); ABC-MCMC algorithm with small threshold (ABC2); adaptive ABC-MCMC algorithm (ABCad1) with adaptation stopped after the burn-in phase; adaptive ABC-MCMC algorithm (ABCad2); marginal sampler (M - M1 - M2).}
\end{figure}

Figure \ref{fig:scen1} (top row) shows the computational time required to perform a single realization, the distance of the latent partition estimate and the true latent partition, and the effective sample size of the entropy of the sampled partitions and the number of clusters, for different sample sizes $n$ and different sampling strategies. The inclusion of an adaptive step is increasing the computational time, which is still increasing slowly as far as the sample size increases. On the counterpart, with the adaptation strategy, the ABC-MCMC algorithm produces estimates of the latent partition which are closer to the true value in terms of VI distance. 
Further, observe that a slight modification of the threshold $\varepsilon$ in the non-adaptive ABC-MCMC algorithm results in drastic changes in terms of performance, hence showing that the adaptive strategy not only produces better estimates, but is also a more stable algorithm. For this reason, from now on, we will only consider the adaptive ABC-MCMC algorithm. 

\subsection{Mixture of univariate g-and-k distributions}\label{sec:gnk_uni}

We then move to a scenario where the density is known up to an intractable constant. We simulate sets of data from an unbalanced mixture of two g-and-k distributions, i.e. 
\[
f_0(x) = 0.75 \psi(x;-3,0.75,-0.9,0.1,0.8) + 0.25 \psi(x;3,0.5,0.4,0.5,0.8)
\]
where $\psi(x;a,b,g,k,c)$ denotes the density function of a g-and-k distribution with location parameter $a$, scale parameter $b$, shape parameter $g$ (mainly affecting the skewness), shape parameter $k$ (mainly affecting the kurtosis), and the parameter $c$ fixed and equal to $0.8$. The g-and-k distribution is then defined through its quantile function $F^{-1}_X(u):[0,1] \to \mathbb R$ with
\begin{equation}
	F^{-1}_X(u) = a + b \left(1 + c\tanh(gu/2)\right)\Phi^{-1}(u)\left(1 + \Phi^{-1}(u)^2\right)^k
\end{equation}
and $\Phi^{-1}(u)$ denotes the quantile function of a standard Gaussian distribution. We consider different sample sizes, with $n \in \{100, 250\}$. 
Our prior model specification consists in a g-and-k mixture with Pitman-Yor process mixing measure. Moreover, the base measure $G_0$ equals a product of independent distribution for the relevant parameters of the model, with $a \sim \mathrm N(0, 25)$, $b \sim \mathrm{IG}(1,2)$, $g \sim \mathrm N(0, 25)$, and $k \sim \mathrm{IG}(1,2)$. 

The performance of the ABC-MCMC sampler are further compared with a marginal sampling scheme with a Monte-Carlo integration to estimate the probability of sampling a new value, in the spirit of Algorithm 8 of \citet{Nea00}, where we consider $m\in\{10,100\}$ temporary values for the Monte-Carlo integration (algorithms M1 and M2 respectively). We remark that while sampling a realization from a g-and-k distribution can be done efficiently, the evaluation of the density require a numerical optimizations, which impacts also the Monte-Carlo integration step. 

Figure \ref{fig:scen1} (bottom row) shows the computational time required to perform a single realization, the distance of the latent partition estimate and the true latent partition, and the effective sample size of the entropy of the sampled partitions and the number of clusters, for different sample sizes $n$ and different sampling strategies.
The ABC-MCMC algorithms are significantly faster than the marginal strategy.
Further, the increased computational cost of the marginal strategy is not translated in more precise estimates of the partition, as shown in Figure~\ref{fig:scen1}. 
The marginal sampler, known for its performances in terms of mixing of the sampled chains, is showing a slightly larger effective sample size but with a larger variance, especially when looking at the Markov chain of the entropy.

\subsection{Mixture of multivariate g-and-k distributions}\label{sec:gnk_multi}

As a multivariate extension of the example in Section~\ref{sec:gnk_uni}, we consider data from the multivariate g-and-k distribution, in dimension $p=2$. The multivariate g-and-k distribution shares the same intractability of univariate one, with the further addition that, to the best of our knowledge, it is not possible to approximate the probability density function numerically. To generate from the bivariate g-and-k distribution it suffices to simulate $(u_1, u_2)$ from a bivariate Gaussian distribution with zero mean, unit marginal variances and correlation $\rho$ and then let 
\[
y_i = a_i + b_i \left(1 + c_i \tanh(g_i u_i / 2)\right) u_i \left(1 + u_i^2\right)^{k_i}, \quad i=1, 2
\]
As in the previous example we assume $c_i$ fixed and equal to $0.8$ and the correlation between the $u_i$'s fixed as $\rho = 0.5$. 
A priori, we assume a multivariate g-and-k mixture model with Pitman-Yor proces mixing measure. The base measure $G_0$ over parameters $\{a_i, b_i, g_i, k_i\} \ i=1, 2$ factorizes into the product of the marginal distributions, that are assumed identical to the ones in Section~\ref{sec:gnk_uni} for each $i=1, 2$. 

\begin{figure}[!h]
	\centering
	\includegraphics[width = 0.99\textwidth]{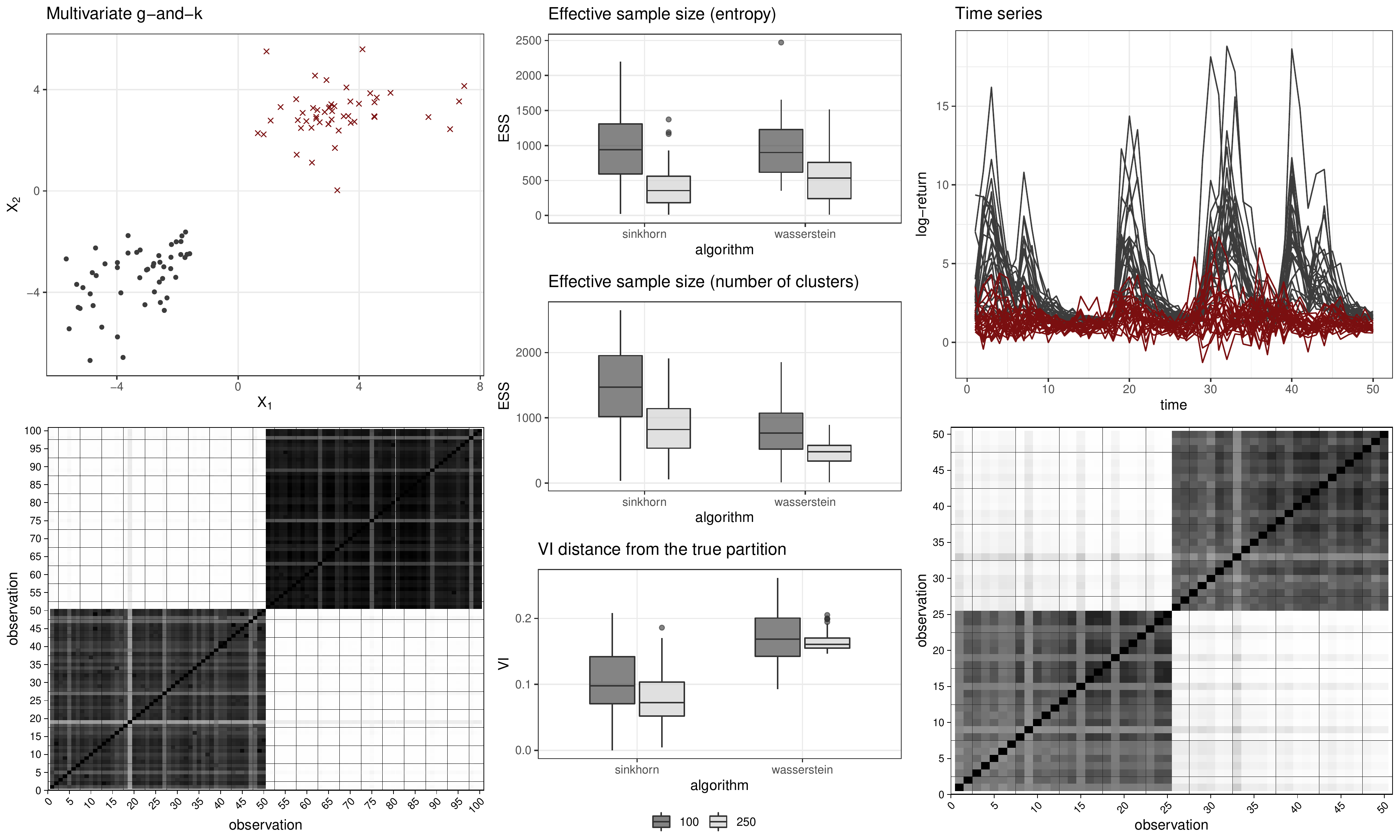}
	\caption{\label{fig:TS} Left column: an example of data generated from the multivariate g-and-k distribution (top) with the posterior estimate of the similarity matrix (bottom). Middle column: effective sample size of the entropy (top), effective sample size of the number of clusters (center), and VI distance from the true partition (bottom) for the multivariate g-and-k scenario. Right column: time series data (top) with the posterior estimate of the similarity matrix (bottom).}
\end{figure}

We simulated data from an equally-weighted mixture of bivariate g-n-k distributions, with parameters (along each direction $i=1,2$) equal to the ones in Section~\ref{sec:gnk_uni}.
The right column of Figure~\ref{fig:TS} shows an example of simulated data (top) with the posterior estimate of the similarity matrix (bottom), while the middle column shows the effective sample size of the entropy (top), the effective sample size for the number of clusters (center), and the VI distance (bottom) evaluated for $100$ of replications, for different sample sizes. It is apparent that both the methods show comparable effective sample sizes, while the Sinkhorn algorithm produces slightly more precise estimates of the latent partitions. As far as the computational cost is concerned, we did not observe any significant difference comparing the runtimes when using the Wasserstein or Sinkhorn distance.

\subsection{Time series stratification}\label{sec:ts}


We consider another scenario where observations are multivariate and the kernel is intractable. 
Specifically, let $y_i =(y_{i,1}, \ldots, y_{i,T}),\  i=1, \ldots, n$  and think of each observation as a time series. 
The kernel $\kernel(\cdot; \theta)$ equals the L\'{e}vy-driven stochastic volatility model \citep{barndorff2002econometric, chopin2013smc2}  with parameters $\theta = (\mu, \beta, \xi, \omega, \lambda)$, see Equation (11) in \cite{Ber19b}. 
The L\'{e}vy-driven stochastic volatility model is popular in financial applications, where it is used to model the log-return of stocks, i.e., given the time series of the prices of the stocks $\{s_{t}\}_{t=1}^{T+1}$, $y_{t} = \log (s_{t+1} - s_t) / s_{t}$. 
A similar scenario, but with a single time series, was analyzed with ABC tools in 
\cite{Ber19b}
as a challenging example in the context of state-space models. Observe how their goal is different from ours: they set to perform inference on the parameters of the stochastic volatility model that generated a single time series, while our is to cluster similar elements belonging to a sample of multiple time series.

As far as the distance between two time series in concerned, we employ the 
\emph{Hilbert distance} on the $1$-lagged time series, introduced in \cite{Ber19b}, see their Section 2.3.2.
Briefly, let us just recall that given a time series $\{y_t\}_{t=1}^T$, the $1$-lagged
time series is the collection of points $\{(y_{t}, y_{t+1})\}_{t=1}^{T-1}$.
Lagged time series, also named delay reconstruction or embeddings, have a key
role in dynamical systems.

We generated $n=50$ time series with $T=50$ observed times, from a two-component mixture with equal weights. In the first component, data ara generated from the L\'{e}vy-driven stochastic volatility model with parameters $(1.5, 2.75, 1.0, 2.5, 1.0)$ while in the second with parameters $(1.0, 2.0, 0.6, 1, 0.4)$, see Figure~\ref{fig:TS} (top-right panel). The base measure $G_0$ is  the product of independent distributions, namely $\mu \sim \mathrm N(1, 4^2)$, $\beta \sim \mathrm{N} (1, 4^2)$, $\xi \sim \mathrm{Ga}(1, 2)$, $\omega \sim \mathrm{Ga}(1, 1)$ and $\lambda \sim \mathrm{Ga}(1, 1)$. 

We ran the adaptive ABC-MCMC (ABCad2) algorithm. The right column of Figure~\ref{fig:TS} shows the posterior similarity matrix (bottom) and the point estimate of the random partition obtained using the greedy algorithm in \cite{Ras18} (top), the adjusted rand index between the estimated and true partition is equal to one. To give a rough estimate of the computational cost, the runtime required by this simulation is approximately three hours on a standard laptop.

\subsection{Clustering a population of networks}\label{sec:graph}

\begin{figure}[h!]
	\begin{subfigure}{.5\textwidth}
		\centering
		\includegraphics[width=.8\linewidth]{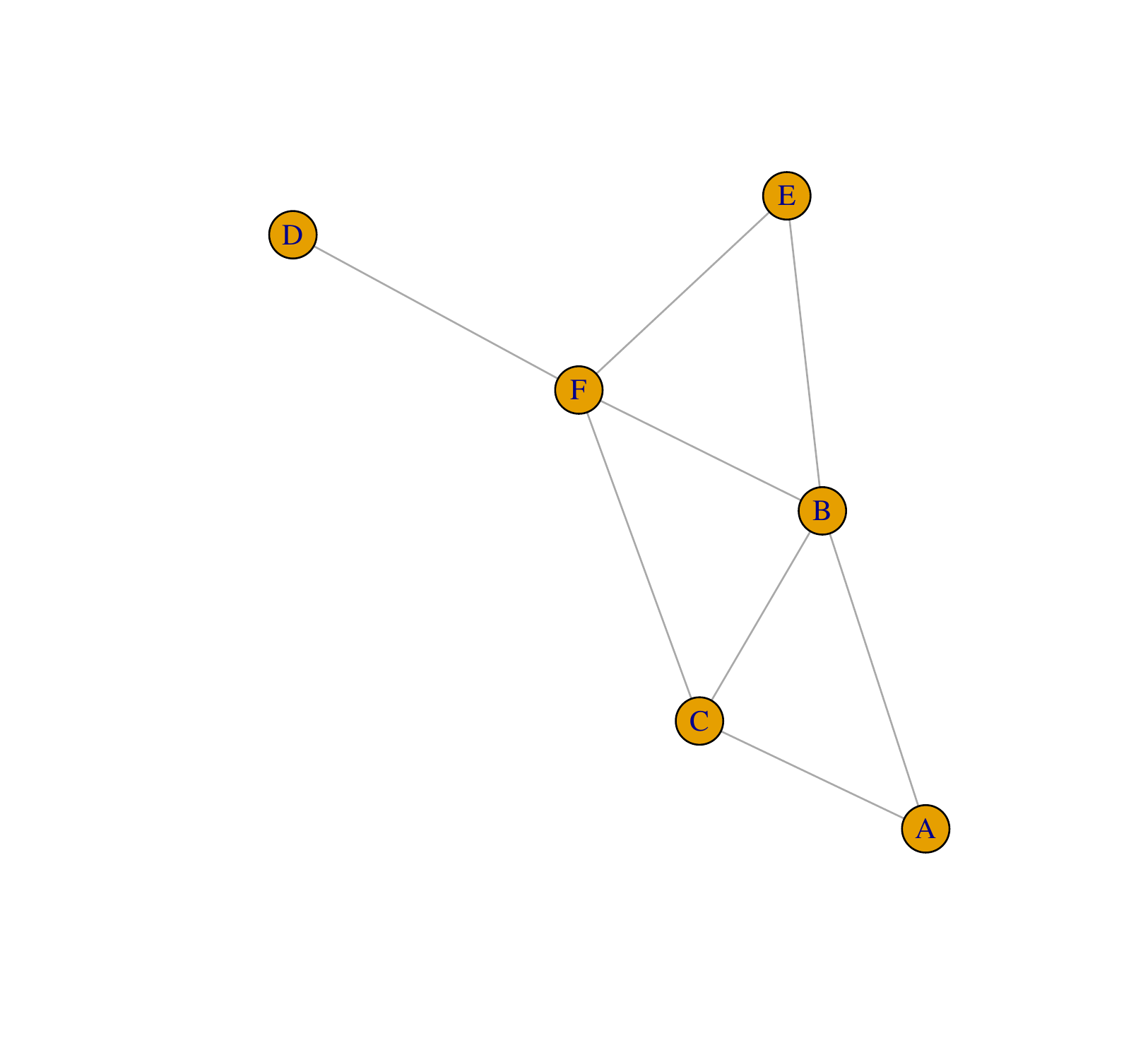}  
	\end{subfigure}
	\begin{subfigure}{.5\textwidth}
		\centering
		\includegraphics[width=.8\linewidth]{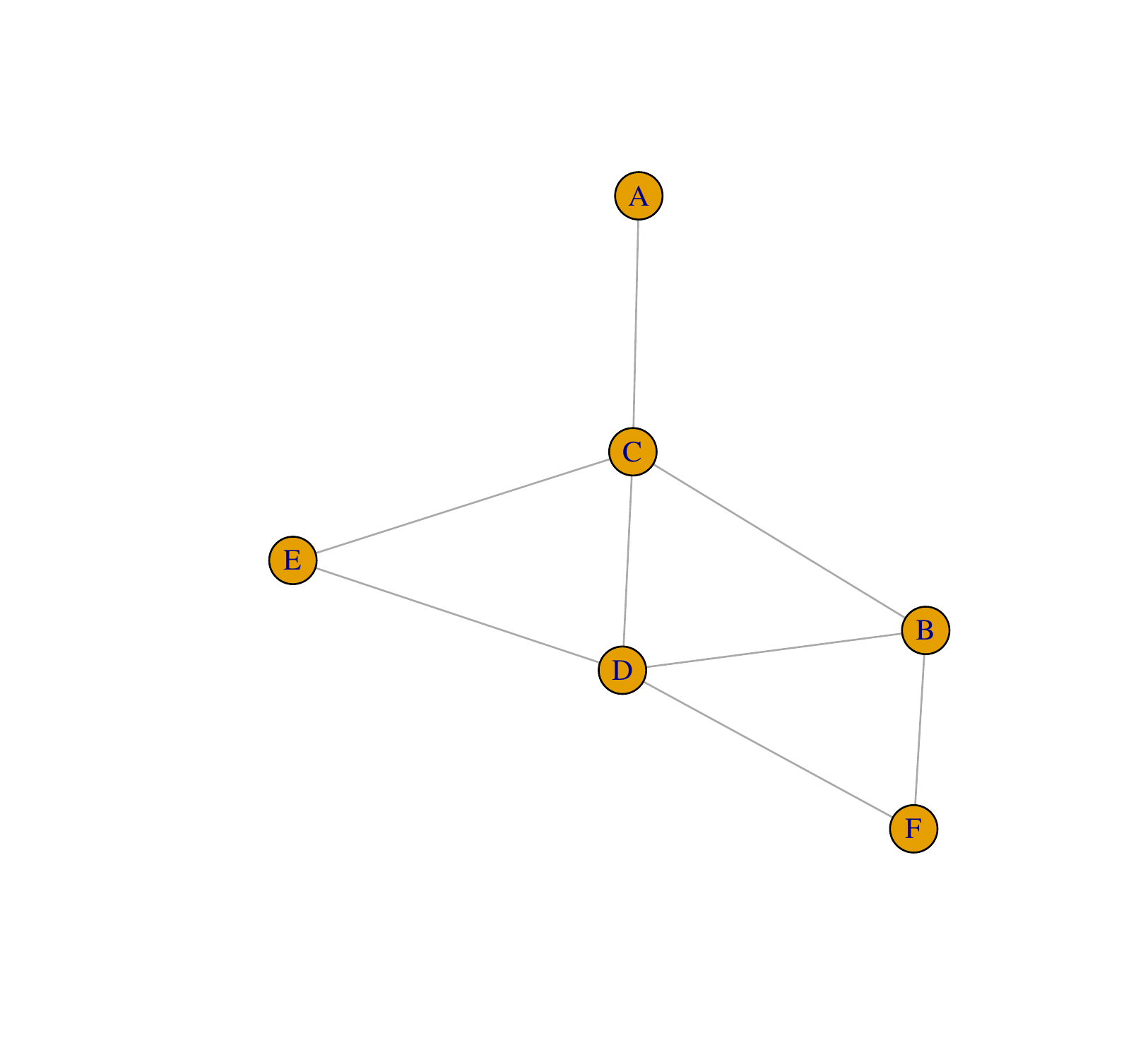}  
	\end{subfigure}
	\caption{An example of two graphs differing only on the labeling of the nodes, but with the same topology.
		The graph in the right picture is recovered starting from the graph in the left picture by renaming the nodes as $D \rightarrow A$, $C \rightarrow B$, $F \rightarrow C$, $B \rightarrow D$, $A \rightarrow F$, $E \rightarrow E$.}
	\label{fig:two_networks}
\end{figure}

As a final illustration, we analyze data from $n=52$ airline companies serving the US airports\footnote{Data are available on the OpenFligths database (\url{https://openflights.org/})}. For each airline, we represent the covered routes as the edges of a graph (also called network) $\bm G_i = \{\mathcal V_i, \mathcal E_i\}$, where $\mathcal V_i$ represent an $M$-dimensional set of nodes (or vertexes) for the observation $i$-th, $i = 1,\dots, n$, and $\mathcal E_i$ denotes the set of tuples $(j,k)\in\mathcal V\times\mathcal V$. 
The airports are shared by all the companies, so that $V_i = V$ for all $i = 1, \dots, n$, and, in particular, the $M=100$ nodes correspond to the $100$ most served airports in the US. We further assume the graph undirected. 
With the aim of clustering together graphs with similar topology, we consider unlabelled networks, so that, for instance, the two networks in Figure \ref{fig:two_networks} are completely identical.

To measure the distance between two specific graphs $\bm G_i$ and $\bm G_j$ we use as cost operator $C(\bm G_i, \bm G_j)$ the spectral distance between graphs, as defined in \citet{gu2015spectral}. Among different possible choices for such metric, this distance is particularly suited for our purpose, as it focuses on the topology of the networks rather than on the labeling of the nodes.

\begin{remark}
	The main difference between this application and the previous sections is that the observed data $\bm G_{1:n}$ are not a subset of $\mathbb{R}^d$ anymore. Nonetheless, the formulation of the Wasserstein distance in Equation \eqref{eq:WEI2}, and then the consequent results, remains valid from general choices of the cost operator $C$.
\end{remark}

We consider as data generating process an Exponential Random Graph Model \citep[ERGM, see, e.g.,][]{Rob07a}.
Recall that we denote by $M$ the number of nodes, assumed fixed, and let $\bm Y$ an $M \times M$ binary matrix such that $Y_{jk} = 0$ if $j$ and $k$ are not connected and $Y_{jk} = 1$ otherwise. In this context, the matrix $\bm Y$ is usually termed adjacency matrix, and it is in one-to-one correspondence  with $\bm G = (\mathcal{V}, \mathcal{E})$, when we assume $\bm G$ an unlabeled network. The assumption underlying ERGMs is that the topology of an observed graph $\bm y$ can be explained by a set of statistics $\bm s(\bm y)$. In particular we assume
\begin{equation}\label{eq:ergm}
	P(\bm Y=\bm y \mid \bm \theta) = \frac{\exp \left(\btheta^\intercal \bm s(y)\right)}{Z_{\btheta}}
\end{equation} 
where $Z_{\btheta}$ is a normalizing constant, not available in closed form. Simulation strategies from \eqref{eq:ergm} are discussed in  \citet{morris2008specification}. 
The model specification is completed by specifying the statistics $\bm s(\bm y)$. Generally the choice of these statistics is problem specific, and there is no one-fits-all choice. For the airlines networks, we have the following structural behaviour 
\begin{enumerate}
	\item[i)] Several networks have a strong hub-and-spoke behavior, meaning that there is one node connected to most of the other ones and, a part from that particular node, the rest of the network is barely connected. This behavior corresponds to companies which have a main airport connected to most of the other ones. 
	\item[ii)] Several networks instead have a more connected topology, meaning that most of the nodes are connected to many other nodes. This behavior represents companies which are diffuse over the airports. 
	\item[iii)] In both cases, there are nodes that are not connected to any other node, i.e. connections not served by the company.
\end{enumerate}
These insight led us to consider statistics of the form: 
\begin{equation*}
	\bm s(\bm y) = \left( \sum_{i, j = 1}^M y_{ij},  \sum_{j=1}^M \uno_{\left[y_{j \bullet}  = 0 \right]},  \sum_{j=1}^M \uno_{\left[y_{j \bullet}  \in [2, 10] \right]},  \sum_{j=1}^M \uno_{\left[y_{j \bullet}  \in [11, 50] \right]}  \right)
\end{equation*}
where $y_{j \bullet} = \sum_{k=1}^M y_{jk}$. Despite the simplicity of the model, maximum likelihood estimates of the parameters $\bm \theta$ for our sample of networks are hard to compute, and most of the time we incur in numerical errors. 
We complete the specification of the model by letting $G_0$ be a four dimensional Gaussian distribution with covariance equal to $10 I$, where $I$ denotes the identity matrix and mean $(-4, 3, 15, -20)$. 
The values of the mean parameters were chosen via an empirical Bayes procedure as the maximum likelihood estimator when considering all the data together.


\begin{figure}[h]
	\centering
	\begin{subfigure}{0.2\linewidth}
		\includegraphics[width=\linewidth]{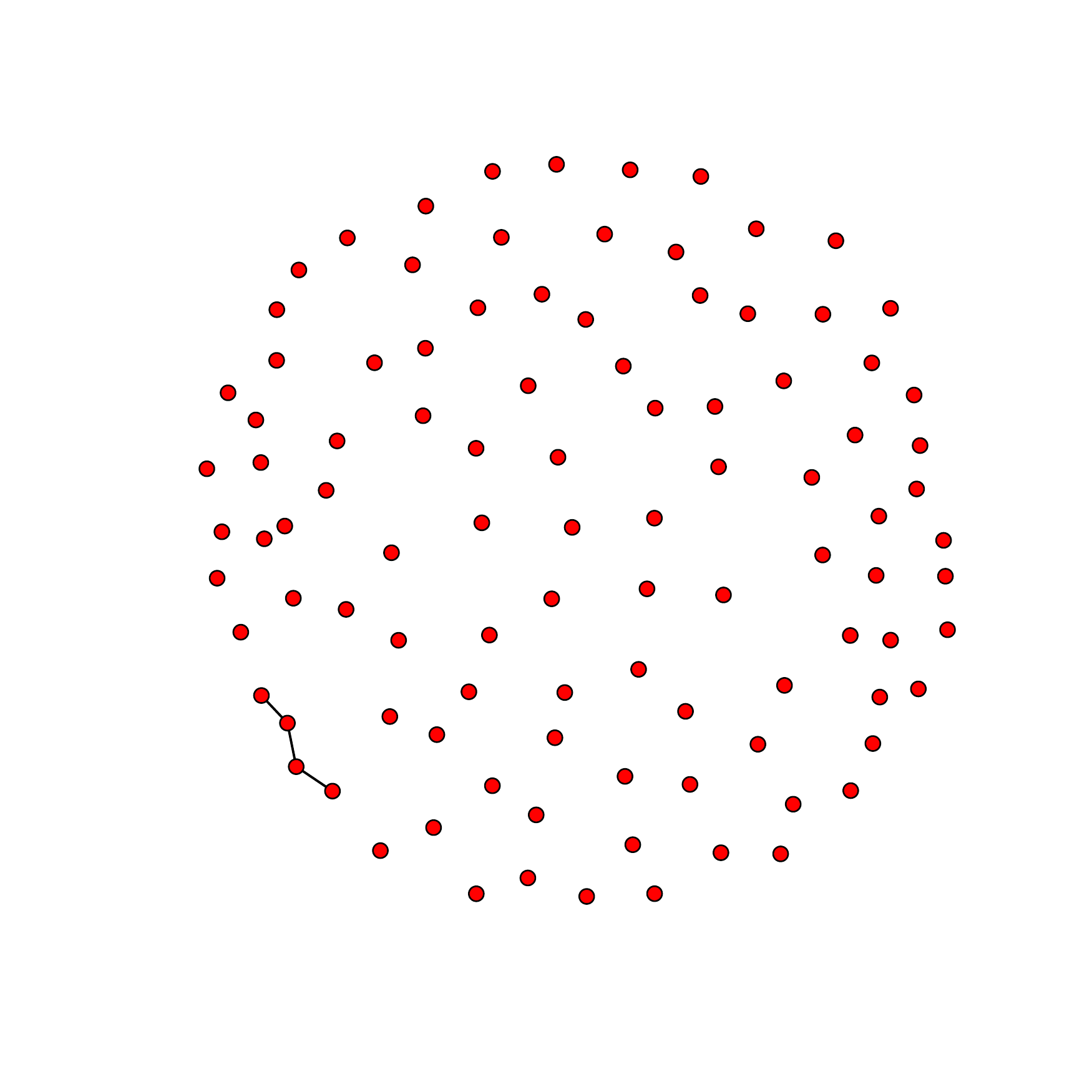}
		\caption{Cluster 1 $(n_1 = 44)$}
	\end{subfigure}\hspace{1cm}%
	\begin{subfigure}{0.2\linewidth}
		\includegraphics[width=\linewidth]{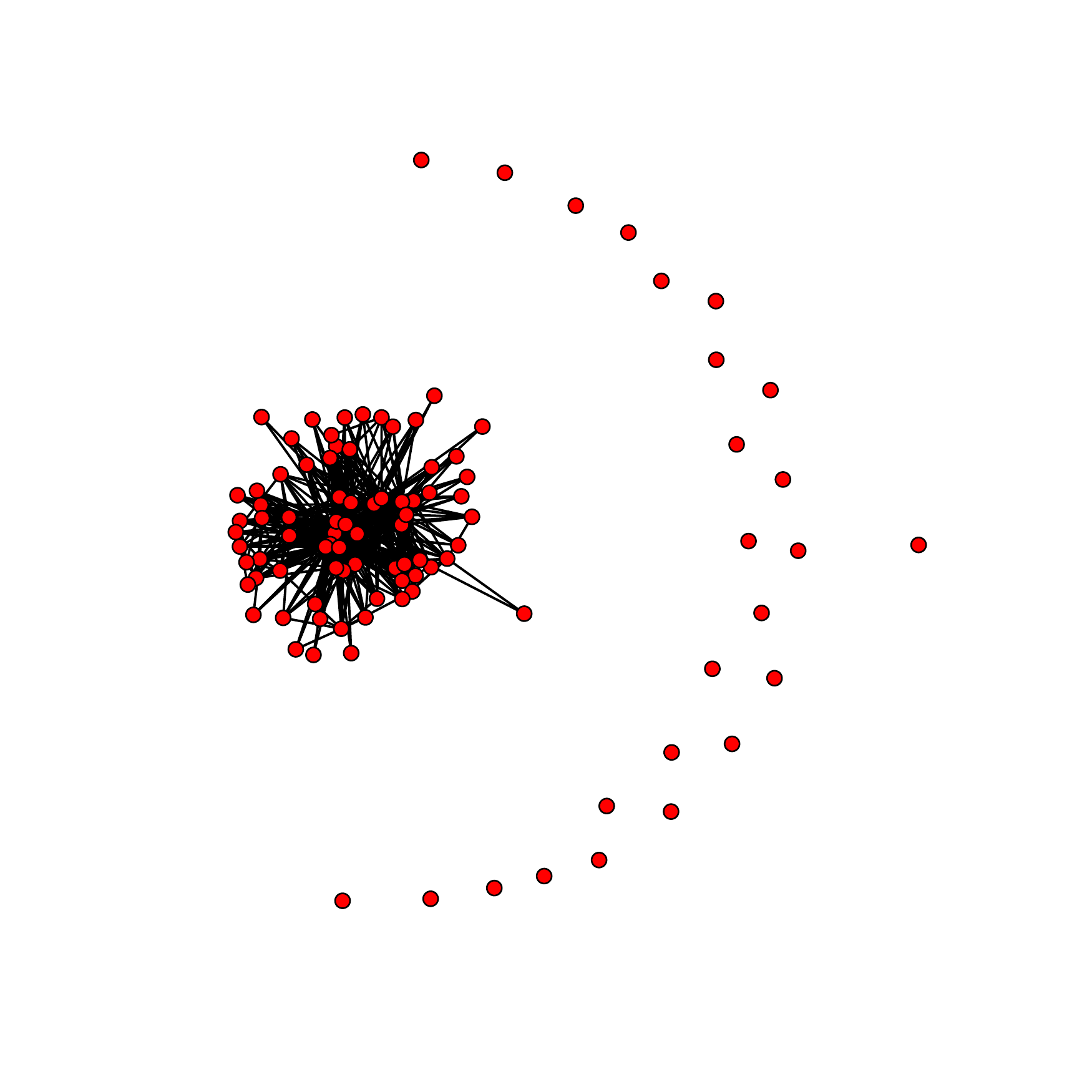}
		\caption{Cluster 2 $(n_2 = 4)$}
	\end{subfigure}\hspace{1cm}%
	\begin{subfigure}{0.2\linewidth}
		\includegraphics[width=\linewidth]{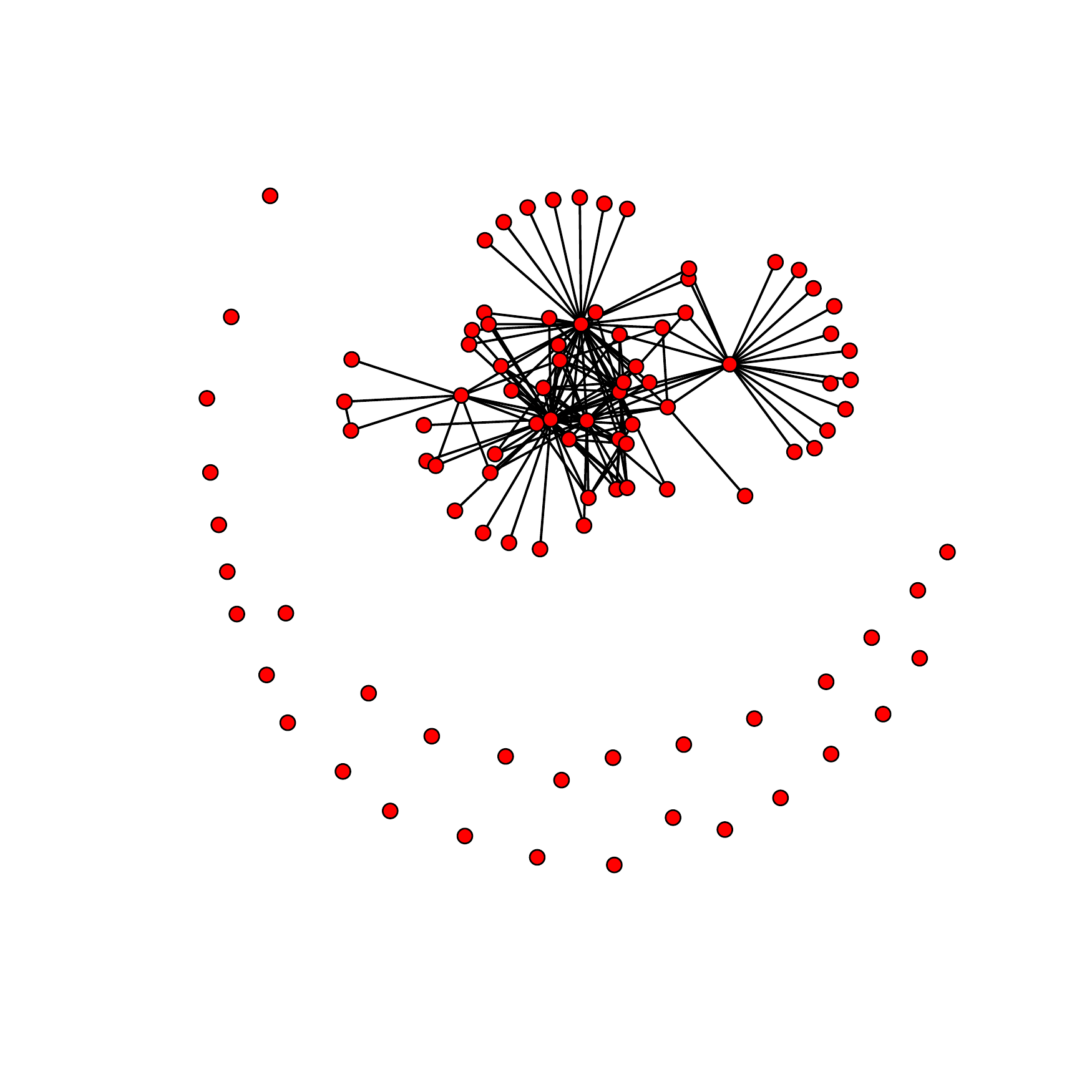}
		\caption{Cluster 3 $(n_3 = 1)$}
	\end{subfigure}
	
	\centering
	\begin{subfigure}{0.2\linewidth}
		\includegraphics[width=\linewidth]{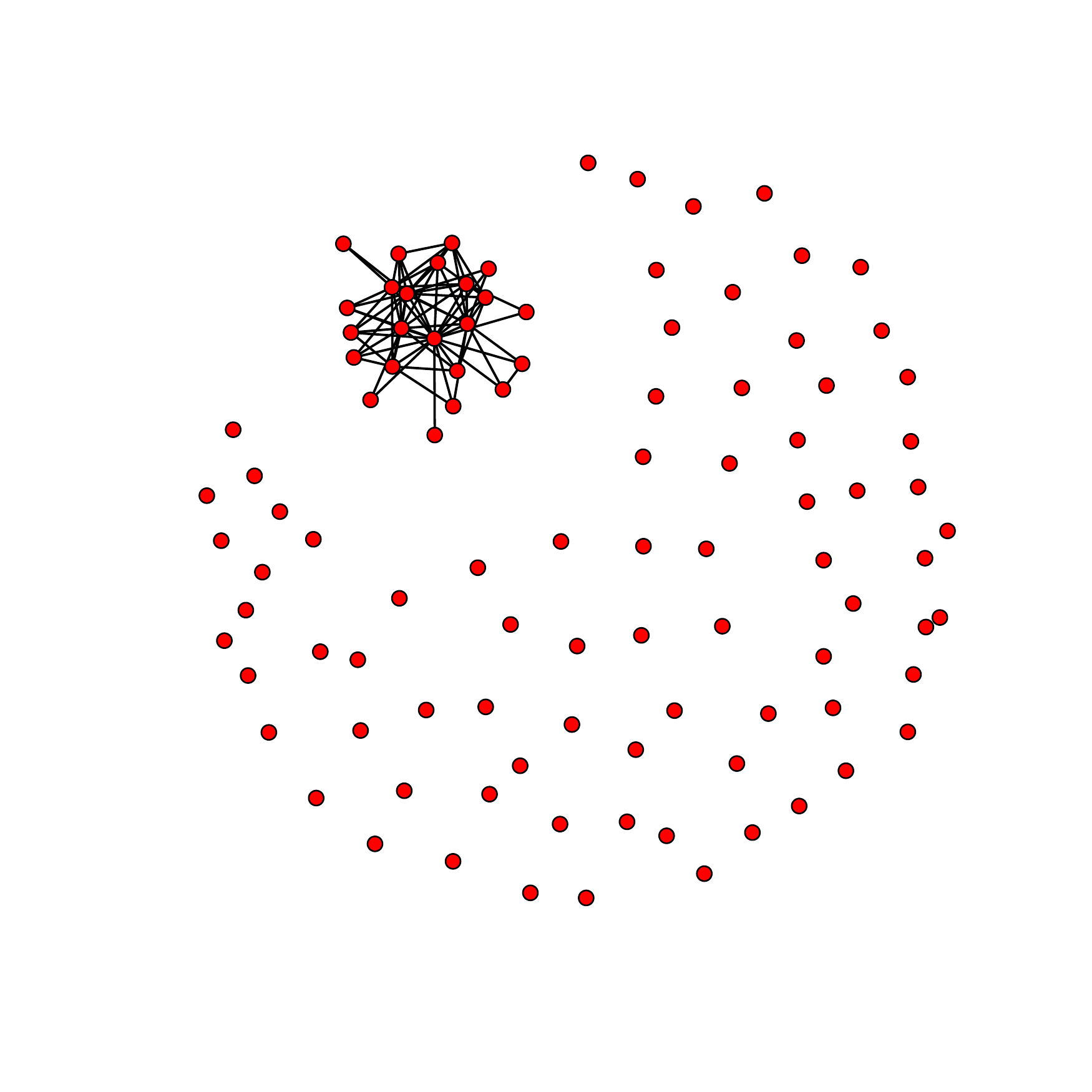}
		\caption{Cluster 4 $(n_4 = 1)$}
	\end{subfigure}\hspace{1cm}%
	\begin{subfigure}{0.2\linewidth}
		\includegraphics[width=\linewidth]{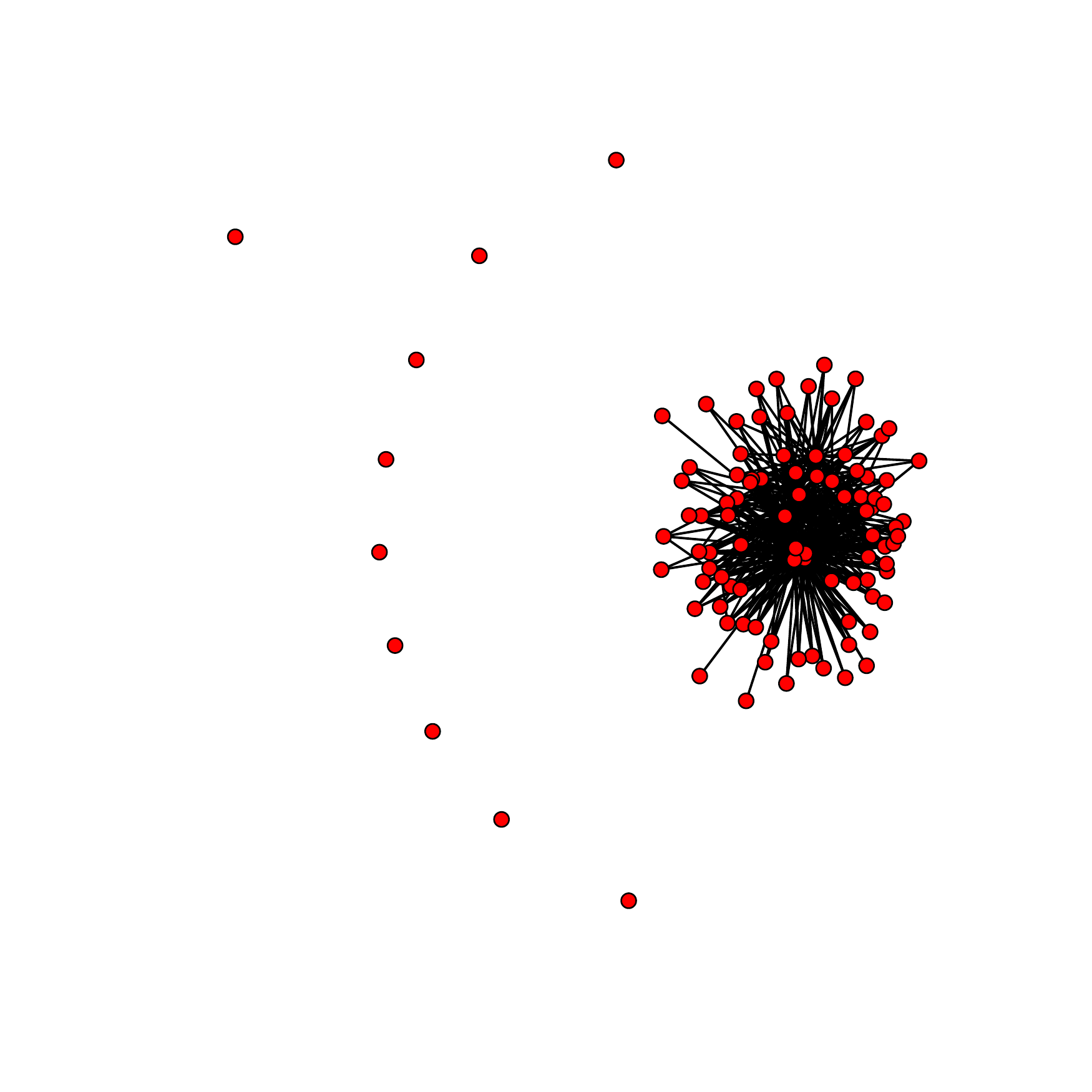}
		\caption{Cluster 5 $(n_5 = 2)$}
	\end{subfigure}
	\caption{Medoids of the 5 clusters of the partition obtained minimizing the Binder loss.}\label{fig:medoids}
\end{figure}

The point estimate of the latent partition identifies 5 clusters, the cluster sizes $n_j$ and the medoid in each cluster are reported in
Figure~\ref{fig:medoids}. Most of the observations belong to a cluster with few connection, see medoid (a) in Figure \ref{fig:medoids}, which corresponds to airlines serving few airports in the network. The clusters corresponding to medoids (b) and (e) are networks where the nodes are highly connected, denoting airlines which are serving several airports. The remaining clusters (c) and (d) are singletons.

\section{Discussion}

In this paper, we introduced an approximate sampling strategy to deal with model based clustering whenever the kernel function is known up to an intractable normalizing constant, but it is easy to define a distance between pairs of observations.
We proposed an ABC-MCMC algorithm, exploiting the predictive distribution induced by the underlying random probability measure, and the use of the Wasserstein distance, in connection to the optimal transportation problems.
Further, we proposed an adaptive strategy to avoid the arduous choice of the threshold $\varepsilon$, 
providing theoretical and numerical results as support.
In extensive simulation studies we have shown that our proposal is a suitable choice in many contexts where the problem is hardly tractable or intractable
despite its simplicity, we have obtained good performance for both computational cost and quality of the estimates, especially for the adaptive extension. The generality of the model allows us to work on abstract spaces, as shown for example in the case of study described in Section~\ref{sec:graph}.

Our algorithm suffers from the curse of dimensionality, as all the other MCMC algorithms for mixture models. In particular, when the dimension of the parameter space increases it becomes more and more difficult to propose suitable values for the cluster parameters, 
while when the dimension of the data increases both the observed and synthetic data suffer from sparsity.
In this situation, we would advise to first project the data on a lower dimensional subspace, via, e.g., principal component analysis, and then perform model-based clustering on the lower dimensional projections.

Several extensions are possible. On the algorithmic side, it could be interesting to add an \textit{acceleration} step to sample the unique values, similar to the algorithms in \cite{Nea00}. This would bring our approach closer to the Gibbs-like algorithm in \cite{Cla20}. 
Moreover, we could consider more general models and, in particular, extensions to the partially exchangeable case.

\section{Acknowledgments}
Mario Beraha gratefully acknowledges the DataCloud laboratory (\url{https://datacloud.polimi.it}); experiments in Sections \ref{sec:gnk_multi}, \ref{sec:ts} and \ref{sec:graph} have been performed thanks to the Cloud resources offered by the DataCloud laboratory. 
Riccardo Corradin gratefully acknowledges the financial support from the Italian Ministry of Education, University and Research (MIUR), “Dipartimenti di Eccellenza" grant 2018-2022, and the DEMS Data Science Lab for supporting this work through computational resources.

\bibliography{MYbib}

\appendix

\section{Proof of Lemma \ref{theo:inv}}\label{app:proof_inv}
Assume $q(\brho_n \to \brho_n^\prime) = \P(\brho_n^\prime \mid \brho_n)$ a genuine proposal as in section \ref{sec:algo}, where we have an acceptance ratio equal to $1$, and the transition kernel is equal to 
\begin{equation}
	r(\brho_n \to \brho_n^\prime) = q(\brho_n \to \brho_n^\prime) \P(\Wei_p(\bm y_{n}, \bm s_{n}) < \varepsilon \mid \brho_n^\prime),
\end{equation}
then we have
\begin{align*}
	\P &(\brho_n \mid \Wei_p(\bm y_{1:n}, \bm s_{1:n}) < \varepsilon) r(\brho_n \to \brho_n^\prime) \\
	&= \P (\brho_n \mid \Wei_p(\bm y_{1:n}, \bm s_{1:n}) < \varepsilon) q(\brho_n \to \brho_n^\prime ) \P (\Wei_p(\bm y_{n}, \bm s_{n})<\varepsilon \mid \brho_n^\prime)\\
	&= \frac{ \P (\brho_n, \Wei_p(\bm y_{n}, \bm s_{n})<\varepsilon)  }{ \P (\Wei_p(\bm y_{n}, \bm s_{n})<\varepsilon) } \frac{\P(\brho_n^\prime, \brho_n)}{\P(\brho_n)} \frac{\P (\Wei_p(\bm y_{n}, \bm s_{n})<\varepsilon , \brho_n^\prime)}{\P (\brho_n^\prime)}\\
	&= \P(\Wei_p(\bm y_{n}, \bm s_{n})<\varepsilon \mid \brho_n) \frac{\P(\brho_n^\prime, \brho_n)}{\P(\brho_n^\prime)} \frac{\P (\Wei_p(\bm y_{n}, \bm s_{n})<\varepsilon , \brho_n^\prime)}{\P (\Wei_p(\bm y_{n}, \bm s_{n})<\varepsilon) }\\
	&= \P(\Wei_p(\bm y_{n}, \bm s_{n})<\varepsilon \mid \brho_n) q( \brho_n^\prime \to \brho_n) \P (\brho_n^\prime \mid \Wei_p(\bm y_{n}, \bm s_{n})<\varepsilon )\\
	&= \P_\varepsilon(\brho_n^\prime \mid \bm y_{1:n}) r(\brho_n^\prime \to \brho_n)
\end{align*}
and the chain respect the detailed balance condition. 

\section{Proof of Theorem \ref{th:adapt}}\label{app:ad_abc_mcmc}
We first check the continuity of the Wasserstein distance between two empirical measures. Recall that from \eqref{eq:WEI2}, we can write the Wasserstein distance as
\[
\Wei_p(\bm y_{1:n}, \bm s_{1:n}) =\left\{ \min_{P \in M_{n\times n} } <C(\bm y_{1:n}, \bm s_{1:n}), P> \right\}^{\frac{1}{q}}
\]
Suppose that the cost matrix is continuous in its elements; this accounts for assuming that the distance is continuous w.r.t the usual $\mathbb{R}^p$ topology, which is always the case for the distances considered in this work. Then we have that for a fixed value of $P$, $<C(\bm y_{1:n}, \bm s_{1:n}), P>$ is continuous, i.e
\[
\lim_{\bdelta \to \bm 0} <C(\bm y_{1:n}, \bm s_{1:n} + \bdelta), P> - <C(\bm y_{1:n}, \bm s_{1:n}), P> = 0 
\]
Hence, the Wasserstein distance is obtained as the minimum of continuous functions, which is itself continuous. We now want to prove that, for $t$ large enough, 
\begin{equation*}
	\P(\brho_n\mid \Wei_p(\bm y_{1:n}, \bm s_{1:n}) < \varepsilon_t) r(\brho_n \to \brho_n^\prime) = 	\P(\brho_n^\prime\mid\Wei_p(\bm y_{1:n}, \bm s_{1:n}) < \varepsilon_t) r(\brho_n^\prime \to \brho_n)\;a.s.
\end{equation*}
Note that 
\begin{align}
	\P(\brho_n\mid &\Wei_p(\bm y_{1:n}, \bm s_{1:n}) < \varepsilon_t) r(\brho_n \to \brho_n^\prime)\nonumber\\
	&=\P(\brho_n\mid \Wei_p(\bm y_{1:n}, \bm s_{1:n}) < \varepsilon_t) q(\brho_n \to \brho_n^\prime) \P(\Wei_p(\bm y_{1:n}, \bm s_{1:n}) < \varepsilon_{t + 1}\mid \brho_n^\prime)\nonumber\\
	&=\frac{\P(\brho_n, \Wei_p(\bm y_{1:n}, \bm s_{1:n}) < \varepsilon_t) }{\P(\Wei_p(\bm y_{1:n}, \bm s_{1:n}) < \varepsilon_t)}\frac{\P(\Pi^\prime_{1:n} , \brho_n)}{\P(\brho_n)} \frac{\P(\Wei_p(\bm y_{1:n}, \bm s_{1:n}) < \varepsilon_{t + 1}, \brho_n^\prime)}{\P( \brho_n^\prime)}\nonumber\\
	&=\frac{\P(\Wei_p(\bm y_{1:n}, \bm s_{1:n}) < \varepsilon_{t}, \brho_n^\prime)}{\P(\Wei_p(\bm y_{1:n}, \bm s_{1:n}) < \varepsilon_t)}
	\frac{\P(\Pi^\prime_{1:n} , \brho_n)}{\P( \brho_n^\prime)}
	\frac{\P(\brho_n, \Wei_p(\bm y_{1:n}, \bm s_{1:n}) < \varepsilon_{t+1}) }{\P(\brho_n)}\nonumber\\
	&\times\left(\frac{\P(\Wei_p(\bm y_{1:n}, \bm s_{1:n}) < \varepsilon_{t+1}, \brho_n^\prime)}{\P(\Wei_p(\bm y_{1:n}, \bm s_{1:n}) < \varepsilon_{t}, \brho_n^\prime)}\right)
	\left(\frac{\P(\brho_n, \Wei_p(\bm y_{1:n}, \bm s_{1:n}) < \varepsilon_{t})}{\P(\brho_n, \Wei_p(\bm y_{1:n}, \bm s_{1:n}) < \varepsilon_{t+1})}\right)\label{eq:lastrow}
\end{align}

Exploiting the first term in \eqref{eq:lastrow}, and by denoting with $p(w\mid\brho_n)$ the density function of $\Wei_p(\bm y_{1:n}, \bm s_{1:n})$, we have
\begin{align*}
	&\lim_{t \to +\infty}\frac{\P(\Wei_p(\bm y_{1:n}, \bm s_{1:n}) < \varepsilon_{t+1}, \brho_n^\prime)}{\P(\Wei_p(\bm y_{1:n}, \bm s_{1:n}) < \varepsilon_{t}, \brho_n^\prime)} =\lim_{t \to +\infty} \frac{\P(\Wei_p(\bm y_{1:n}, \bm s_{1:n}) < \varepsilon_{t+1}\mid \brho_n^\prime)}{\P(\Wei_p(\bm y_{1:n}, \bm s_{1:n}) < \varepsilon_{t}\mid \brho_n^\prime)}\\
	&\lim_{t \to +\infty}=\frac{\int_0^{\varepsilon_{t+1}}p(\d w\mid\brho_n)}{\int_0^{\varepsilon_{t}}p(\d w\mid\brho_n)} = 1 +  \lim_{t \to +\infty}\frac{\int_{\varepsilon_t}^{\varepsilon_{t+1}}p(\d w\mid\brho_n)}{\int_0^{\varepsilon_{t}}p(\d w\mid\brho_n)} \stackrel{a.s.}{=} 1
\end{align*}
where the last equation holds thanks to the fact that $\lim_{t \to +\infty}\lvert\epsilon_{t+1} - \epsilon_{t}\rvert = 0$, and due to assumption A1 in Lemma \ref{th:adapt}. 
With similar consideration, we have that the second term in \eqref{eq:lastrow} is a.s. equal to $1$. In force of that, we have 
\begin{align*}
	\P(\brho_n\mid &\Wei_p(\bm y_{1:n}, \bm s_{1:n}) < \varepsilon_t) r(\brho_n \to \brho_n^\prime)\\
	&\stackrel{a.s.}{=}\frac{\P(\Wei_p(\bm y_{1:n}, \bm s_{1:n}) < \varepsilon_{t}, \brho_n^\prime)}{\P(\Wei_p(\bm y_{1:n}, \bm s_{1:n}) < \varepsilon_t)}
	\frac{\P(\Pi^\prime_{1:n} , \brho_n)}{\P( \brho_n^\prime)}
	\frac{\P(\brho_n, \Wei_p(\bm y_{1:n}, \bm s_{1:n}) < \varepsilon_{t+1}) }{\P(\brho_n)}\\
	&= \P(\brho_n^\prime\mid \Wei_p(\bm y_{1:n}, \bm s_{1:n}) < \varepsilon_{t}) q(\brho_n^\prime \to \brho_n)\P(\Wei_p(\bm y_{1:n}, \bm s_{1:n}) < \varepsilon_{t+1} \mid \brho_n)\\
	&= \P(\brho_n^\prime\mid \Wei_p(\bm y_{1:n}, \bm s_{1:n}) < \varepsilon_{t}) r(\brho_n^\prime \to \brho_n)
\end{align*}
We proved that, for $t\to +\infty$, the invariant distribution is $P(\brho_n = \brho_n \mid \Wei_p(\bm y_{1:n}, \bm s_{1:n}) < \varepsilon_t)$ a.s., and due to $\varepsilon_{t}\to\varepsilon^*$, we have that  $P(\brho_n = \brho_n \mid \Wei_p(\bm y_{1:n}, \bm s_{1:n}) < \varepsilon^*)$ a.s., which concludes the proof.
\end{document}